\documentclass[runningheads]{llncs}
\usepackage{stpa}
\usepackage{version}
\usepackage[hidelinks]{hyperref}
\hypersetup{colorlinks=true,linkcolor=blue,citecolor=blue,urlcolor=blue}
\usepackage{orcidlink}

\title
 {Space-Time Process Algebra \\ with Asynchronous Communication}
\author
 {J.A. Bergstra\,\orcidlink{0000-0003-2492-506X} \and 
  C.A. Middelburg\,\orcidlink{0000-0002-8725-0197}}
\institute
 {Informatics Institute, Faculty of Science, University of Amsterdam \\
  Science Park~900, 1098~XH Amsterdam, the Netherlands \\
  \email{J.A.Bergstra@uva.nl, C.A.Middelburg@uva.nl}}

\titlerunning
 {Space-Time Process Algebra with Asynchronous Communication}
\authorrunning
 {J.A. Bergstra \and C.A. Middelburg}

\begin{document}

\maketitle

\begin{abstract}
%% 114 %%
We introduce a process algebra that concerns the timed behaviour of 
distributed systems with a known spatial distribution. 
This process algebra provides a communication mechanism that deals with 
the fact that a datum sent at one point in space can only be received at 
another point in space at the point in time that the datum reaches that 
point in space. 
The variable-binding integration operator used in related process 
algebras to model such a communication mechanism is absent from this 
process algebra.
This is considered an advantage because the variable-binding operator 
does not really fit in with an algebraic approach and a process algebra 
with this operator is not firmly founded in established metatheory.
\begin{classcode}
C.2.4, D.2.1, D.2.4, F.1.2, F.3.1.
% MSC 2000: 68M14, 68N30, 68Q60, 68Q85.
\end{classcode}
\end{abstract}

\section{Introduction}
\label{sect-introduction}

In~\cite{BB93a}, a generalization of the process algebra known as
\ACP\ (Algebra of Communicating Processes)~\cite{BK84b} is introduced 
that concerns the timed behaviour of distributed systems with a known 
spatial distribution.
Communication within such a system is unavoidably asynchronous because 
it takes time to transmit data from one point in space to another point 
in space.
In~\cite{BM03b}, this process algebra is adapted to a setting with 
urgent actions.
In such a setting, which is justified in e.g.~\cite{Mid02b}, it is 
possible for two or more actions to be performed consecutively at the 
same point in time.
In~\cite{BB93a}, as well as in~\cite{BM03b}, it was demonstrated that
the process algebra introduced could be used to describe simple
protocols transmitting data via an intermediate station that moves in
space.
To model a mechanism for asynchronous communication in space-time, both 
process algebras provide the so-called integration operator.

The integration operator is a variable-binding operator.
The fact that an algebraic theory with a variable-binding operator is 
not fully algebraic may be considered a minor issue in itself.
There is, however, another issue that is largely a consequence of this 
fact.
All extensions and generalizations of \ACP\ without variable-binding 
operators are firmly founded in established meta-theory from the fields
of universal algebra and structural operational semantics.
Many definitions of well-known notions from these fields need to be 
generalized for variable-binding operators and consequently well-known 
results need no longer hold in the presence of variable-binding 
operators.
The issue is that extensions and generalizations of \ACP\ with 
variable-binding operators, unlike those without variable-binding 
operators, are not firmly founded in established meta-theory.

This issue was the main reason to start the work presented in this 
paper, namely the devising of a process algebra that concerns the timed 
behaviour of systems with a known spatial distribution and that provides 
a mechanism for asynchronous communication in space-time that does not 
involve variable-binding operators.
In the process algebra devised, called \STPA\ (Space-Time Process 
Algebra), the operators that model the asynchronous communication 
mechanism are state operators (see~\cite{BW90}) of a special kind.
They are reminiscent of the operators from~\cite{BKT85} that model other 
kinds of asynchronous communication, kinds that are primarily used in 
cases where synchronous communication is an option as well.

Except for its adaptation to a setting with urgent actions, \STPA\ is 
closer to the process algebra introduced in~\cite{BB93a} than to the one 
introduced in~\cite{BM03b}.
The most striking difference with both earlier process algebras is of
course the absence of the integration operator.
There are two additional important differences between \STPA\ and the 
process algebra introduced in~\cite{BB93a}.
Firstly, a different approach has been followed in \STPA\ to allow for 
absolute timing and relative timing to be mixed.
Both kinds of timing have been put on the same footing and consequently
the variable-binding initial abstraction operator could be disposed of.
Secondly, two auxiliary operators used in~\cite{BB93a} to axiomatize
parallel composition, viz.\ the bounded initialization operator and the
ultimate delay operator, have been replaced in \STPA\ by one operator 
for the only combination in which they are used in~\cite{BB93a}.
The replacing operator is called the time-out operator.

In the case of asynchronous communication in space-time, reception of a
datum is usually given priority over idling.
This calls for special priority operators, known as the maximal progress 
operators (cf.~\cite{BB93a,BM03b}).
Therefore, \STPA\ is extended with the appropriate priority operators.
The resulting process algebra is called \STPAt.
A closed term over the signature of \STPA\ or \STPAt\ denotes a process 
with a finite upper bound to the number of actions that it can perform. 
Guarded recursion allows the description of processes without a finite 
upper bound to the number of actions that it can perform.
Therefore, \STPA\ and \STPAt\ are extended with guarded recursion.

\STPA\ and \STPAt, extended with guarded recursion, are primarily 
intended for the detailed description of the timed behavior of 
distributed systems with a known spatial distribution. 
Important examples of such systems are data communication protocols 
where the data are transmitted through space.
The work presented in this paper also includes the description of a 
simple example of such a protocol.

In \STPA, \STPAt, and their extensions with guarded recursion, the 
mathematical structure for points in time, periods of time, and 
coordinates of points in space is the signed meadow with square root 
whose domain is the set of real numbers.
This structure has a purely equational axiomatization.
Signed meadows and signed meadows with square root are introduced 
in~\cite{BBP13a}.

Because CCS (Calculus of Communicating Systems)~\cite{Mil80,Mil89} and 
ACP are closely related, it should be relatively easy to devise a 
CCS-based variant of \STPA.
It is perhaps more difficult to devise a variant of \STPA\ based on 
CSP (Communicating Sequential Processes)~\cite{BHR84,Hoa85} because in 
CSP equality corresponds to failure equivalence instead of bisimilarity.

This paper is organized as follows.
First, we give a brief summary of signed meadows with square root
(Section~\ref{sect-MD}).
Next, we make some introductory remarks on \STPA\ 
(Section~\ref{sect-intro-STPA}),
present and informally explain the constants and operators of \STPA\ 
(Section~\ref{sect-sign-STPA}), and 
present and discuss the axioms of \STPA\ 
(Section~\ref{sect-axioms-STPA}).
After that, we extend \STPA\ with maximal progress 
(Section~\ref{sect-maxpr}) and guarded recursion 
(Section~\ref{sect-recursion}).
Following this, an example of the use of \STPAt\ with guarded recursion 
is given (Section~\ref{sect-example}).
Thereafter, we give a structural operational semantics for \STPA, 
\STPAt, and their extensions with guarded recursion and define a 
notion of bisimilarity based on this (Section~\ref{sect-SOS}). 
Then, we present soundness and (semi-)completeness results with 
respect to bisimilarity for the axioms of \STPA\ with guarded recursion 
and \STPAt\ with guarded recursion (Section~\ref{sect-sound-complete}).
Finally, we make some concluding remarks 
(Section~\ref{sect-conclusions}).

\section{The Signed Meadow of Reals with Square Root}
\label{sect-MD}

In the process algebra introduced in this paper, the mathematical 
structure for points in time, periods of time, and coordinates of points 
in space is the signed meadow with square root whose domain is the set 
of real numbers, shortly called the \emph{signed meadow of reals with 
square root}.
In this section, we give a brief summary of the signature and equational
theory of this structure.

A meadow is a field with the multiplicative inverse operation made total
by imposing that the multiplicative inverse of zero is zero.
A signed meadow is a meadow expanded with the signum (or sign) operation.
By the presence of the signum operation, the ordering $<$ on the domain 
of a signed meadow that corresponds to the usual ordering becomes 
definable (see below).
A signed meadow with square root is a signed meadow expanded with a 
square root operation that is made total by imposing that the square 
root of an element of the domain is the additive inverse of the square 
root of the additive inverse of the element if the element is less than 
zero.
The reasons for choosing this structure are that it is appropriate and 
it has a purely equational axiomatization.
Signed meadows and signed meadows with square root originate 
from~\cite{BBP13a}.

The signature of signed meadows with square root consists of the 
following constants and operators:
\begin{itemize}
\item
the constants $0$ and $1$;
\item
the binary \emph{addition} operator
${} + {}$;
\item
the binary \emph{multiplication} operator
${} \mul {}$;
\item
the unary \emph{additive inverse} operator
$\mathop{-} {}$;
\item
the unary \emph{multiplicative inverse} operator
${} \minv$;
\item
the unary \emph{signum} operator
$\sign$;
\item
the unary \emph{square root} operator
$\sqrt{\phantom{u}}$.
\end{itemize}
The constants and operators from this signature are adopted from real 
arithmetic, which gives an appropriate intuition about them.
Because the signum operator is perhaps not widely known, we mention that 
the signum of a real number is $1$, $0$, or $-1$ according to whether 
the number is greater than, equal to, or less than $0$.

We assume that there is a countably infinite set $\cU$ of variables, 
which contains $u$, $v$, and $w$.
Terms are built as usual.
We use infix, prefix, and postfix notation as usual.
We use the usual precedence convention to reduce the need for
parentheses.

A signed meadow with square root is an algebra with the signature of 
signed meadows with square root that satisfies the equations given in 
Table~\ref{eqns-meadow}.%
\begin{table}[!t]
\caption{Axioms for signed meadows with square root}
\label{eqns-meadow}
\begin{eqntbl}
\begin{eqncol}
(u + v) + w = u + (v + w)                                  \\
u + v = v + u                                              \\
u + 0 = u                                                  \\
u + (-u) = 0                                               \\
(u \mul v) \mul w = u \mul (v \mul w)                      \\
u \mul v = v \mul u                                        \\
u \mul 1 = u                                               \\
u \mul (v + w) = u \mul v + u \mul w                       \\
{(u\minv)}\minv = u                                        \\
u \mul (u \mul u\minv) = u
\end{eqncol}
\qquad \qquad
\begin{eqncol}
\sign(u \divi u) = u \divi u                               \\
\sign(1 - u \divi u) = 1 - u \divi u                       \\
\sign(-1) = -1                                             \\
\sign(u\minv) = \sign(u)                                   \\
\sign(u \mul v) = \sign(u) \mul \sign(v)                   \\
(1 - \frac{\sign(u) - \sign(v)}{\sign(u) - \sign(v)}) \mul
(\sign(u + v) - \sign(u)) = 0
\\ {} \\
\sqrt{u\minv} = (\sqrt{u})\minv                            \\
\sqrt{u \mul v} = \sqrt{u} \mul \sqrt{v}                   \\
\sqrt{u^2 \mul \sign(u)} = u                               \\
\sign(\sqrt{u} - \sqrt{v}) = \sign(u - v)
\end{eqncol}
\end{eqntbl}
\end{table}
From these equations, among others, the equations $0\minv = 0$ and 
$\sqrt{u} = - \sqrt{- u}$ can be derived.
The relatively involved sixth equation on the right-hand side of 
Table~\ref{eqns-meadow} tells us that the conditional equation 
$\sign(u) = \sign(v) \Implies \sign(u + v) = \sign(u)$ 
holds in a signed meadow with square root.
In~\cite{BBP15a}, it is shown that an equation of terms over the 
signature of signed meadows is derivable from the equations given in 
Table~\ref{eqns-meadow} iff it holds in the signed meadow of reals.

In signed meadows with square root, the \emph{subtraction} operation 
$-$ the \emph{division} operation $/$\,, and the \emph{squaring} 
operation ${}^2$ are defined as follows:
\pagebreak[2]
\begin{ldispl}
\begin{aeqns}
u - v     & = & u + (-v)\;, \\
u \divi v & = & u \mul v\minv\;, \\
u^2       & = & u \mul u\;,
\end{aeqns} 
\end{ldispl}%
the \emph{less than} predicate $<$ and the \emph{less than or equal} 
predicate $\leq$ are defined as follows:
\begin{ldispl}
\begin{aeqns}
u < v    & \Iff & \sign(u - v) = -1\;,
\\
u \leq v & \Iff & \sign(\sign(u - v) - 1) = -1\;,
\end{aeqns}
\end{ldispl}%
and the \emph{minimum} and \emph{maximum} operations $\min$ and $\max$ 
are defined as follows:
\begin{ldispl}
\displaystyle
\begin{aeqns}
\min(u,v) & = &
{\displaystyle \frac{\sign(\sign(u - v) - 1)}{\sign(\sign(u - v) - 1)}}
  \mul (u - v) + v\;,
\\[1.25em]
\max(u,v) & = &
{\displaystyle \frac{\sign(\sign(u - v) + 1)}{\sign(\sign(u - v) + 1)}}
  \mul (u - v) + v\;.
\end{aeqns}
\end{ldispl}%

\section{Introductory Remarks on Space-Time Process Algebra}
\label{sect-intro-STPA}

\STPA\ (Space-Time Process Algebra) is an \ACP-style process algebra 
with timed and spatially located actions that provides a mechanism for 
asynchronous communication in space-time.
Before the constants and operators of \STPA\ are presented and 
informally explained in Section~\ref{sect-sign-STPA}, some introductory 
remarks on \STPA\ are in order.

For simplicity, it is assumed in \STPA\ that data are transmitted with 
velocity $v$ through space in all directions and can be detected at any 
distance. 
It should not be difficult to take into account issues such as signal 
strength degradation and receivers threshold.
If a process sends a datum at a point in space $\xi$ and a point in time 
$t$, then that datum can be received by another process at a point in 
space $\xi'$ and a point in time $t'$ provided that the distance between 
$\xi$ and $\xi'$ is $v \cdot (t' - t)$.

In \STPA, a distinction is made between potential and actual send and 
receive actions.
An action that is potentially capable of sending a datum at a point in 
space $\xi$ and a point in time $t$ may become actually capable of doing 
so if $t$ is no later than the current point in time.
An action that is potentially capable of receiving a datum at a point in 
space $\xi$ and any point in time between $t$ and $t'$ may become 
actually capable of doing so if $t'$ is no later than the current point 
in time and the datum reaches the point in space $\xi$ at a point in 
time between $t$ and $t'$.

It is important that the communication mechanism of \STPA\ takes into 
account the fact that a datum sent at one point in space can only be 
received at another point in space at the point in time that the datum 
reaches the latter point in space.
Whether a process that is potentially capable of sending a given datum
at a given point in space and a given point in time is actually capable 
of doing so depends only on the current point in time.
However, whether a process potentially capable of receiving a given 
datum at a given point in space and a given point in time is actually 
capable of doing so depends on the current point in time, the previously 
sent data, and the points in space and points in time at which these 
data were sent.

The communication mechanism of \STPA\ is modelled by state operators.
The state that is involved comprises, for each sending of a datum that
has taken place, the datum concerned, the point in space at which it was 
sent, and the point in time at which it was sent.
The state is updated when a datum is sent.
The state is not updated when a datum is received, because it may later 
be received at a point in space further away from the point in space 
from which it was sent.
From the state, the set of future points in times at which a given datum 
can be received at a given point in space can be determined.
Notice that, because a sent datum may be received more than once,
the asynchronous communication mechanism modelled by the state operators
of \STPA\ is of a broadcasting nature.

The state operators of \STPA\ differ from the state operators used 
in~\cite{BB93a,BM03b} to model asynchronous communication mainly in that 
they involve both state and point in time in their effect.
They are `initialization and actualization' operators in the sense that 
they make a process start at a certain point in time and then actualize 
potential send and receive actions of the process. 

To keep it simple, some details have been omitted from the above 
introductory remarks.
First of all, the remarks are made from an absolute timing point of 
view. 
However, because it can be convenient, relative timing is also provided 
for potential send and receive actions.
Moreover, it is assumed that communication takes place via channels.
Channels can be considered abstractions of the frequency bands at which 
data is transmitted.

\section{Space-Time Process Algebra: Constants and Operators}
\label{sect-sign-STPA}

This section presents and informally explains the constants and 
operators of \STPA.
The axioms of \STPA\ are presented in Section~\ref{sect-axioms-STPA}.

In \STPA, it is assumed that a fixed but arbitrary finite set $\Chan$ of 
\emph{channels} and a fixed but arbitrary finite set $\Data$ of 
\emph{data} have been given.
The elements of $\Time$ are taken as points in time and the
elements of $\Place$ are taken as points in space.
 
To keep track of all sendings of a datum that have taken place, the 
communication mechanism of \STPA\ makes use of communication states. 

The set $\State$ of \emph{communication states} is defined as follows:%
\footnote
{We write $\fpset(S)$, where $S$ is a set, for the set of all finite
 subsets of $S$.}
\begin{ldispl}
\State = \fpset(\Chan \x \Data \x \Time \x \Place)\;.
\end{ldispl}%

Let $\sigma$ be a communication state.
Then $(c,d,t,\xi) \in \sigma$ indicates that in communication state 
$\sigma$ the sending of the datum $d$ via the channel $c$ at the point 
in space $\xi$ and the point in time $t$ has taken place.

Below, we present the signature of \STPA.
Shortly therafter, a brief informal explanation of the constants and 
operators from the signature of \STPA\ is given.

The signature of \STPA\ consists of the following constants and 
operators:
\begin{itemize}
\item
the \emph{immediate inaction} constant $\dead$;
\item
the \emph{absolutely timed inaction} constant $\dead(t)$ \\ for each 
$t \in \Time \union \set{\infty}$;
\item
the \emph{relatively timed inaction} constant $\dead[t]$ \\ for each 
$t \in \Time \union \set{\infty}$;
\item
the \emph{absolutely timed potential send action} constant 
$\apsnd{c}{d}{t}{\xi}$ \\ for each $c \in \Chan$, $d \in \Data$, 
$t \in \Time$, and $\xi \in \Place$;
\item
the \emph{relatively timed potential send action} constant 
$\rpsnd{c}{d}{t}{\xi}$ \\ for each $c \in \Chan$, $d \in \Data$, 
$t \in \Time$, and $\xi \in \Place$;
\item
the \emph{absolutely timed potential receive action} constant
$\aprcv{c}{d}{t}{t'}{\xi}$ \\ for each $c \in \Chan$, $d \in \Data$, 
$t \in \Time$ and $t' \in \Time \union \set{\infty}$ with $t < t'$, 
and $\xi \in \Place$;
\item
the \emph{relatively timed potential receive action} constant
$\rprcv{c}{d}{t}{t'}{\xi}$ \\ for each $c \in \Chan$, $d \in \Data$, 
$t \in \Time$ and $t' \in \Time \union \set{\infty}$ with $t < t'$, 
and $\xi \in \Place$;
\item
the \emph{absolutely timed actual send action} constant
$\aesnd{c}{d}{t}{\xi}$ \\ for each $c \in \Chan$, $d \in \Data$, 
$t \in \Time$, and $\xi \in \Place$;
\item
the \emph{absolutely timed actual receive action} constant
$\aercv{c}{d}{t}{\xi}$ \\ for each $c \in \Chan$, $d \in \Data$, 
$t \in \Time$, and $\xi \in \Place$;
\item
the binary \emph{alternative composition} or \emph{choice} operator 
${} \altc {}$;
\item
the binary \emph{sequential composition} operator ${} \seqc {}$;
\item
the binary \emph{parallel composition} or \emph{merge} operator 
${} \parc {}$;
\item
the binary \emph{left merge} operator ${} \leftm {}$;
\item
the binary \emph{time-out} operator $ {} \tout {}$;
\item
the unary \emph{state} operator $\state{C}{t}{\sigma}$ \\ for each 
$C \subseteq \Chan$, $t \in \Time$, and $\sigma \in \State$.
\end{itemize}

We assume that there is a countably infinite set $\cX$ of variables 
which contains $x$, $y$, and $z$, with and without subscripts.
Terms over the signature of \STPA\ are built as usual.
The set $\Proc$ of \emph{process terms} is the set of all closed terms 
over the signature of \STPA.

We use infix notation for the binary operators.
Moreover, we use the following precedence conventions to reduce the need 
for parentheses: the operator $\altc$ binds weaker than all other binary
operators and the operator $\seqc$ binds stronger than all other binary
operators.

Let $c \in \Chan$, $d \in \Data$, $t \in \Time$ and 
$t' \in \Time \union \set{\infty}$ with $t < t'$, $\xi \in \Place$, 
$\sigma \in \State$, $P,Q \in \Proc$, and $C \subseteq \Chan$.
Intuitively, the constants and operators introduced above can be explained 
as follows:
\begin{itemize}
\item
$\dead$ is not capable of doing anything;
\item
$\adead{t}$ is capable of idling till the point in time $t$ and after
that it is not capable of doing anything;%
\footnote
{Throughout the paper, ``till'' stands for ``up to and not including''.}
\item
$\rdead{t}$ is only capable of idling for the period of time $t$ and 
after that it is not capable of doing anything;
\item
$\apsnd{c}{d}{t}{\xi}$ is potentially capable of idling till the point 
in time $t$ and sending the datum $d$ via the channel $c$ at the point 
in space $\xi$ and the point in time $t$ and next terminating 
successfully;
\item
$\rpsnd{c}{d}{t}{\xi}$ is potentially capable of idling for the period 
of time $t$ and sending the datum $d$ via the channel $c$ at the point 
in space $\xi$ after the period of time $t$ and next terminating 
successfully;
\item
$\aprcv{c}{d}{t}{t'}{\xi}$ is potentially capable of idling till a 
point in time $t''$ between $t$ and $t'$ and receiving the datum $d$ via 
the channel $c$ at the point in space $\xi$ and the point in time $t''$ 
and next terminating successfully;
\item
$\rprcv{c}{d}{t}{t'}{\xi}$ is potentially capable of idling for a period 
of time $t''$ between $t$ and $t'$ and receiving the datum $d$ via the
channel $c$ at the point in space $\xi$ after the period of time $t''$ 
and next terminating successfully;
\item
$\aesnd{c}{d}{t}{\xi}$ is actually capable of idling till the point in 
time $t$ and sending the datum $d$ via the channel $c$ at the point in 
space $\xi$ and the point in time $t$ and next terminating successfully;
\item
$\aercv{c}{d}{t}{\xi}$ is actually capable of idling till the point in 
time $t$ and receiving the datum $d$ via the channel $c$ at the point in 
space $\xi$ and the point in time $t$ and next terminating successfully;
\item
$P \altc Q$ behaves either as $P$ or as $Q$, but not both;
\item
$P \seqc Q$ first behaves as $P$ and $Q$ in sequence;
\item
$P \parc Q$ behaves as $P$ and $Q$ in parallel;
\item
$P \leftm Q$ behaves the same as $P \parc Q$, except that it starts
with performing a step of $P$;
\item
$P \tout Q$ behaves the same as $P$, except that it is restricted to
perform its first step not later than the ultimate point in time till
which $Q$ can idle;
\item
$\state{C}{t}{\sigma}(P)$ behaves as $P$ placed in an environment where 
the communication mechanism of \STPA\ is in force for communication via 
the channels in $C$, started at the point in time $t$ from the
communication state $\sigma$.
\end{itemize}

The constant $\dead$ and the operators ${} \altc {}$, ${} \seqc {}$, and 
${}\parc {}$ are well-known in process algebra.
However, when it comes to timed behavior, some remarks about idling may
be appropriate for the operators ${} \altc {}$ and ${}\parc {}$.

In $P_1 \altc P_2$, there is an arbitrary choice between $P_1$ and 
$P_2$.
The choice is resolved on one of them performing its first action, and 
not otherwise.
Consequently, the choice between two idling processes will always be
postponed until at least one of the processes can perform its first
action.
Only when both processes cannot idle any longer, further postponement
is not an option.
If the choice has not yet been resolved when one of the processes 
cannot idle any longer, the choice will simply not be resolved in its 
favour. 

In $P_1 \parc P_2$, $P_1$ and $P_2$ are merged as follows: first either 
$P_1$ or $P_2$ performs its first step and next it proceeds in 
parallel with the process following that step and the process that did 
not perform an step.
However, $P_1$ and $P_2$ may have to idle before they can perform their
first step.
Therefore, $P_1 \parc P_2$ can only start with performing an step of 
$P_1$ or $P_2$ if it can do so before or at the ultimate point of time 
for the other process to start performing steps or to deadlock.

There are constants of \STPA\ in which a point in time or period of time 
is~$\infty$.
The easiest way to deal with that is to extend the predicates $<$ and 
$\leq$ on $\Time$ to $\Time \union \set{\infty}$ as follows:
\begin{ldispl}
\begin{tabular}[t]{@{}l@{}}
$t < \infty$ and $\infty \not< t$ 
for all $t \in \Time$,
\\
$t \leq \infty$ and $\infty \not\leq t$ 
for all $t \in \Time$,
\\
$\infty \not< \infty$ and $\infty \leq \infty$.
\end{tabular}
\end{ldispl}

In the coming sections, there is a need to refer to different sets of 
actions.

The sets
$\PAct$ of \emph{potential actions}, 
$\AAct$ of \emph{actual actions}, 
$\ATAct$ of \emph{absolutely timed actions}, 
$\RTAct$ of \emph{relatively timed actions}, and
$\PRAct$ of \emph{potential receive actions}
are defined as follows:
\begin{ldispl}
\begin{aeqns}
\PAct   & = & \APS \union \APR \union \RPS \union \RPR\;, \\
\AAct   & = & \AAS \union \AAR\;, \\
\ATAct  & = & \APS \union \APR \union \AAS \union \AAR\;, \\
\RTAct  & = & \RPS \union \RPR\;, \\
\PRAct  & = & \APR \union \RPR\;, 
\end{aeqns} 
\end{ldispl} 
where
\begin{ldispl}
\begin{aeqns}
\APS  & = & 
\set{\apsnd{c}{d}{t}{\xi} \where c \in \Chan, d \in \Data, 
                                 t \in \Time, \xi \in \Place}\;, \\
\APR  & = & 
\set{\aprcv{c}{d}{t}{t'}{\xi} \where c \in \Chan, d \in \Data,
                  t \in \Time, t' \in \Time \union \set{\infty},
                                  \xi \in \Place \And t < t'}\;, \\
\RPS  & = & 
\set{\rpsnd{c}{d}{t}{\xi} \where c \in \Chan, d \in \Data,
                                 t \in \Time, \xi \in \Place}\;, \\
\RPR  & = & 
\set{\rprcv{c}{d}{t}{t'}{\xi} \where c \in \Chan, d \in \Data,
                  t \in \Time, t' \in \Time \union \set{\infty}, 
                                  \xi \in \Place \And t < t'}\;, \\
\AAS  & = & 
\set{\aesnd{c}{d}{t}{\xi} \where c \in \Chan, d \in \Data,
                                 t \in \Time, \xi \in \Place}\;, \\
\AAR  & = & 
\set{\aercv{c}{d}{t}{\xi} \where c \in \Chan, d \in \Data,
                                 t \in \Time, \xi \in \Place}\;.
\end{aeqns} 
\end{ldispl} 

The set $\Proc$ of process terms includes terms that are considered to 
denote atomic processes.

The set $\AProc$ of \emph{atomic process terms} is the smallest set 
satisfying the following rules:
\begin{itemize}
\item
if $a \in \PAct \union \AAct$, then $a \in \AProc$;
\item
if $t \in \Time$, then $\adead{t}, \rdead{t} \in \AProc$;
\item
if $\alpha \in \AProc$ and $P \in \Proc$, then 
$\alpha \tout P \in \AProc$.
\end{itemize}

From the above definitions it follows directly that 
$\PAct \union \AAct = \ATAct \union \RTAct$ and 
$\PAct \union \AAct \subset \AProc$.
It may seem strange that a term of the form $\alpha \tout P$, where
$\alpha \in \AProc$ but $P \in \Proc$, is considered to denote an atomic
process.
Recall, however, that the second operand of ${} \tout {}$ is only used 
to restrict the points in time at which the first operand can perform 
its first action.

\section{Space-Time Process Algebra: Axiom System}
\label{sect-axioms-STPA}

In this section, the axiom system of \STPA\ is presented and discussed.
Many axioms of \STPA\ are axiom schemas with a side-condition.
The earliest-time, latest-time, channel, and reception-time-set 
functions referred to in the side-conditions are defined first.

Recall that in \STPA\ the mathematical structure for points in time, 
periods of time, and coordinates of points in space is the signed meadow 
of reals with square root reviewed in Section~\ref{sect-MD}.

In the axiomatization of the time-out operator, we use the 
\emph{earliest-time} function $\lbt$ from $\PAct \union \AAct$ to 
$\Time$ and the \emph{latest-time} function $\ubt$ from
$\PAct \union \AAct$ to $\Time \union \set{\infty}$ defined below.

For each $a \in \PAct \union \AAct$, $\lbt(a)$ is the unique 
$t \in \Time$ such that, for some $c \in C$, $d \in \Data$, and 
$\xi \in \Place$, one of the following holds:
\begin{itemize}
\setlength{\itemsep}{.25ex}
\item
$a \in \set{\apsnd{c}{d}{t}{\xi}, \rpsnd{c}{d}{t}{\xi},
            \aesnd{c}{d}{t}{\xi}, \aercv{c}{d}{t}{\xi}}$;
\item
for some $t' \in \Time \union \set{\infty}$ with $t < t'$,
$a \in \set{\aprcv{c}{d}{t}{t'}{\xi}, \rprcv{c}{d}{t}{t'}{\xi}}$.
\end{itemize}
\pagebreak[2]
For each $a \in \PAct \union \AAct$, $\ubt(a)$ is the unique 
$t' \in \Time \union \set{\infty}$ such that, for some $c \in C$, 
$d \in \Data$, and $\xi \in \Place$, one of the following holds: 
\begin{itemize}
\setlength{\itemsep}{.25ex}
\item
$a \in \set{\apsnd{c}{d}{t}{\xi}, \rpsnd{c}{d}{t}{\xi},
            \aesnd{c}{d}{t}{\xi}, \aercv{c}{d}{t}{\xi}}$;
\item
for some $t \in \Time$ with $t < t'$,
$a \in \set{\aprcv{c}{d}{t}{t'}{\xi}, \rprcv{c}{d}{t}{t'}{\xi}}$.
\end{itemize}

Clearly, $\lbt(a) \neq \ubt(a)$ only if $a$ is an (absolutely or 
relatively timed) potential receive action.
This means that $\lbt(a) = \ubt(a)$ if 
$a \notin \PRAct$.
To emphasize this, we write $\bt(a)$ instead of $\lbt(a)$ or $\ubt(a)$ 
if $a \notin \PRAct$.

In the axiomatization of the state operators, we use the \emph{channel} 
function $\chan$ from $\PAct \union \AAct$ to $\Chan$ defined below.

For each $a \in \PAct \union \AAct$, $\chan(a)$ is the unique 
$c \in \Chan$ such that, for some $d \in \Data$, $t \in \Time$, and 
$\xi \in \Place$, one of the following holds: 
\begin{itemize}
\setlength{\itemsep}{.25ex}
\item
$a \in \set{\apsnd{c}{d}{t}{\xi}, \rpsnd{c}{d}{t}{\xi},
            \aesnd{c}{d}{t}{\xi}, \aercv{c}{d}{t}{\xi}}$;
\item
for some $t' \in \Time \union \set{\infty}$ with $t < t'$,
$a \in \set{\aprcv{c}{d}{t}{t'}{\xi}, \rprcv{c}{d}{t}{t'}{\xi}}$.
\end{itemize}

In the axiomatization of the state operators, we also use the 
\emph{reception-time-set} function $\rcpt$ from 
$\State \x \Chan \x \Data \x \Time \x (\Time \union \set{\infty}) \x
 \Place$ 
to $\fpset(\Time)$ defined below.
 
For all $\sigma \in \State$, $c \in \Chan$, $d \in \Data$, 
$t \in \Time$, $t' \in \Time \union \set{\infty}$, and $\xi \in \Place$,
$\rcpt(\sigma,c,d,t,t',\xi)$ is the set of all $s \in \Time$ with
$t \leq s \leq t'$ such that
\begin{ldispl}
\Exists{s' \in \Time} 
 {(s' \leq t' \And  
   \Exists{\xi' \in \Place} 
    {((c,d,s',\xi') \in \sigma \And (s - s') \mul v = \dist(\xi,\xi')
     )})}\;,
\end{ldispl}%
where $v \in \Time$ is the transmission speed of data and $\dist$ is the 
\emph{distance} function  from $\Place \x \Place$ to $\Dist$, which is 
defined as usual:
\begin{ldispl}
\dist((u_1,v_1,w_1),(u_2,v_2,w_2)) =
\sqrt{(u_2 - u_1)^2 + (v_2 - v_1)^2 + (w_2 - w_1)^2}\;.
\end{ldispl}%

According to the above definition of $\rcpt$, a point in time $s$ 
belongs to $\rcpt(\sigma,c,d,t,t',\xi)$ if $s$ lies between $t$ and $t'$ 
and, according to the communication state $\sigma$, the datum $d$ was 
sent via channel $c$ at a point in time $s'$ before $t'$ and a point in 
space $\xi'$ at a distance $(s - s') \cdot v$ from $\xi$.
Intuitively, the function $\rcpt$ can be explained as follows: 
$\rcpt(\sigma,c,d,t,t',\xi)$ is the set of points in time between $t$ 
and $t'$ at which datum $d$ can be received via channel $c$ at point in 
space $\xi$ in the case where the communication state is $\sigma$.
Notice that $\rcpt(\sigma,c,d,t,t',\xi) = \emptyset$ iff there are no 
points in time between $t$ and $t'$ at which datum $d$ can be received 
via channel $c$ at point in space $\xi$ in the case where the 
communication state is $\sigma$, and that 
$\min(\rcpt(\sigma,c,d,t,t',\xi))$ is the earliest point in time between 
$t$ and $t'$ at which datum $d$ can be received via channel $c$ at point 
in space $\xi$ in the case where the communication state is~$\sigma$.

\nopagebreak[2]
The axiom system of \STPA\ consists of the equations given in 
Tables~\ref{eqns-STPA-I} and~\ref{eqns-STPA-II}.%
\begin{table}[p]
\caption{Axioms of \STPA\ (Part I)}
\label{eqns-STPA-I}
\begin{eqntbl}
\begin{ceqncol}
x \altc y = y \altc x                                                 \\
(x \altc y) \altc z = x \altc (y \altc z)                             \\
x \altc x = x                                                         \\
(x \altc y) \seqc z = x \seqc z \altc y \seqc z                       \\
(x \seqc y) \seqc z = x \seqc (y \seqc z)                             \\
x \altc \dead = x                                                     \\
\dead \seqc x = \dead                                                 \\
x \parc y = x \leftm y \altc y \leftm x                               \\
\alpha \leftm x = (\alpha \tout x) \seqc x                            \\
\alpha \seqc x \leftm y = (\alpha \tout y) \seqc (x \parc y)          \\
(x \altc y) \leftm z = x \leftm z \altc y \leftm z                    \\
\adead{t} \altc \adead{t'} = \adead{t}  & \mif t' < t                 \\
\adead{t} \altc \adead{t'} = \adead{t'} & \mif t \leq t'              \\
a \altc \adead{t} = a 
       & \mif a \notin \PRAct \And a \in \ATAct \And \bt(a) = t \\
\adead{t} \seqc x = \adead{t}                                         \\
\rdead{t} \altc \rdead{t'} = \rdead{t}  & \mif t' < t                 \\
\rdead{t} \altc \rdead{t'} = \rdead{t'} & \mif t \leq t'              \\
a \altc \rdead{t} = a 
       & \mif a \notin \PRAct \And a \in \RTAct \And \bt(a) = t \\
\rdead{t} \seqc x = \rdead{t}                                         \\
\dead = \rdead{0}                                                     \\
\adead{t} \tout \adead{t'} = \adead{t'} & \mif t' < t                 \\
\adead{t} \tout \adead{t'} = \adead{t}  & \mif t \leq t'              \\
\rdead{t} \tout \rdead{t'} = \rdead{t'} & \mif t' < t                 \\
\rdead{t} \tout \rdead{t'} = \rdead{t}  & \mif t \leq t'              \\
a \tout a' = a & \mif \ubt(a) \leq \lbt(a')                 \\
x \tout a = x \tout \adead{t}
       & \mif a \notin \PRAct \And a \in \ATAct \And \bt(a) = t \\
x \tout a = x \tout \rdead{t}      
       & \mif a \notin \PRAct \And a \in \RTAct \And \bt(a) = t \\
x \tout (y \altc z) = x \tout y \altc x \tout z                       \\
x \tout y \seqc z = x \tout y                                         \\
x \tout (y \tout z) = (x \tout y) \tout z                             \\
a \tout \adead{t} = \adead{t}
       & \mif a \in \ATAct \And t < \lbt(a)                       \\
a \tout \adead{t} = a
       & \mif a \in \ATAct \And \ubt(a) \leq t                    \\
a \tout \rdead{t} = \rdead{t}
       & \mif a \in \RTAct \And t < \lbt(a)                       \\
a \tout \rdead{t} = a
       & \mif a \in \RTAct \And \ubt(a) \leq t                    \\
(x \altc y) \tout z = x \tout z \altc y \tout z                       \\
x \seqc y \tout z = (x \tout z) \seqc y                               \\
(x \tout y) \tout z = (x \tout z) \tout y
\end{ceqncol}
\end{eqntbl}
\end{table}
\begin{table}[!p]
\caption{Axioms of \STPA\ (Part II)}
\label{eqns-STPA-II}
\begin{eqntbl}
\begin{ceqncol}
\state{C}{t}{\sigma}(\adead{t'}) = \adead{t}         & \mif t' < t    \\
\state{C}{t}{\sigma}(\adead{t'}) = \adead{t'}        & \mif t \leq t' \\
\state{C}{t}{\sigma}(\rdead{t'}) = \adead{t+t'}                       \\
\state{C}{t}{\sigma}(a) = a                  & \mif \chan(a) \notin C \\
\state{C}{t}{\sigma}(\apsnd{c}{d}{t'}{\xi}) = \adead{t}
                                                     & \mif t' < t    \\
\state{C}{t}{\sigma}(\apsnd{c}{d}{t'}{\xi}) = \aesnd{c}{d}{t'}{\xi} 
                                                     & \mif t \leq t' \\
\state{C}{t}{\sigma}(\rpsnd{c}{d}{t'}{\xi}) = \aesnd{c}{d}{t+t'}{\xi} \\
\state{C}{t}{\sigma}(\aprcv{c}{d}{t'}{t''}{\xi}) = \adead{t} 
              & \mif t'' \leq t \\
\state{C}{t}{\sigma}(\aprcv{c}{d}{t'}{t''}{\xi}) = \adead{t''}
              & \mif t < t'' \And
                     \rcpt(\sigma,c,d,\max(t,t'),t'',\xi) = \emptyset \\
\state{C}{t}{\sigma}(\aprcv{c}{d}{t'}{t''}{\xi}) =
\aercv{c}{d}{t'''}{\xi} 
   & \mif t < t'' \And
          \rcpt(\sigma,c,d,\max(t,t'),t'',\xi) \neq \emptyset \And {} \\
   & \phantom{\mif} 
          t''' = \min(\rcpt(\sigma,c,d,\max(t,t'),t'',\xi))           \\
\state{C}{t}{\sigma}(\rprcv{c}{d}{t'}{t''}{\xi}) = \adead{t+t''} 
     & \mif \rcpt(\sigma,c,d,t+t',t+t'',\xi) = \emptyset              \\
\state{C}{t}{\sigma}(\rprcv{c}{d}{t'}{t''}{\xi}) =
\aercv{c}{d}{t'''}{\xi}       
     & \mif \rcpt(\sigma,c,d,t+t',t+t'',\xi) \neq \emptyset \And {}   \\
     & \phantom{\mif} t''' = \min(\rcpt(\sigma,c,d,t+t',t+t'',\xi))   \\
\state{C}{t}{\sigma}(\aesnd{c}{d}{t'}{\xi}) = \adead{t} & \mif t' < t \\
\state{C}{t}{\sigma}(\aesnd{c}{d}{t'}{\xi}) = \aesnd{c}{d}{t'}{\xi}
                                                     & \mif t \leq t' \\
\state{C}{t}{\sigma}(\aercv{c}{d}{t'}{\xi}) = \adead{t} & \mif t' < t \\
\state{C}{t}{\sigma}(\aercv{c}{d}{t'}{\xi}) = \aercv{c}{d}{t'}{\xi}
                                                     & \mif t \leq t' \\
\state{C}{t}{\sigma}(a \seqc x) = a \seqc \state{C}{t}{\sigma}(x)
                                             & \mif \chan(a) \notin C \\
\state{C}{t}{\sigma}(\apsnd{c}{d}{t'}{\xi} \seqc x) = \adead{t}
     & \mif t' < t                                                    \\
\state{C}{t}{\sigma}(\apsnd{c}{d}{t'}{\xi} \seqc x) = 
\aesnd{c}{d}{t'}{\xi} \seqc \state{C}{t'}{\sigma'}(x)
     & \mif t \leq t' \And \sigma' = \sigma \union \set{(c,d,t',\xi)} \\
\state{C}{t}{\sigma}(\rpsnd{c}{d}{t'}{\xi} \seqc x) =
\aesnd{c}{d}{t+t'}{\xi} \seqc \state{C}{t+t'}{\sigma'}(x) 
     & \mif \sigma' = \sigma \union \set{(c,d,t+t',\xi)}              \\
\state{C}{t}{\sigma}(\aprcv{c}{d}{t'}{t''}{\xi} \seqc x) = \adead{t}
     & \mif t'' \leq t                                                \\
\state{C}{t}{\sigma}(\aprcv{c}{d}{t'}{t''}{\xi} \seqc x) = \adead{t''}
     & \mif t < t'' \And
            \rcpt(\sigma,c,d,\max(t,t'),t'',\xi) = \emptyset          \\
\state{C}{t}{\sigma}(\aprcv{c}{d}{t'}{t''}{\xi} \seqc x) = 
\aercv{c}{d}{t'''}{\xi} \seqc \state{C}{t'''}{\sigma}(x)
   & \mif t < t'' \And
          \rcpt(\sigma,c,d,\max(t,t'),t'',\xi) \neq \emptyset \And {} \\
   & \phantom{\mif} 
          t''' = \min(\rcpt(\sigma,c,d,\max(t,t'),t'',\xi))           \\
\state{C}{t}{\sigma}(\rprcv{c}{d}{t'}{t''}{\xi} \seqc x) = \adead{t+t''} 
     & \mif \rcpt(\sigma,c,d,t+t',t+t'',\xi) = \emptyset              \\
\state{C}{t}{\sigma}(\rprcv{c}{d}{t'}{t''}{\xi} \seqc x) = 
\aercv{c}{d}{t'''}{\xi} \seqc \state{C}{t'''}{\sigma}(x)
     & \mif \rcpt(\sigma,c,d,t+t',t+t'',\xi) \neq \emptyset \And {}   \\
     & \phantom{\mif} t''' = \min(\rcpt(\sigma,c,d,t+t',t+t'',\xi))   \\
\state{C}{t}{\sigma}(\aesnd{c}{d}{t'}{\xi} \seqc x) = \adead{t}
                                                     & \mif t' < t    \\
\state{C}{t}{\sigma}(\aesnd{c}{d}{t'}{\xi} \seqc x) = 
\aesnd{c}{d}{t'}{\xi} \seqc \state{C}{t'}{\sigma}(x) & \mif t \leq t' \\
\state{C}{t}{\sigma}(\aercv{c}{d}{t'}{\xi} \seqc x) = \adead{t}
                                                     & \mif t' < t    \\
\state{C}{t}{\sigma}(\aercv{c}{d}{t'}{\xi} \seqc x) = 
\aercv{c}{d}{t'}{\xi} \seqc \state{C}{t'}{\sigma}(x) & \mif t \leq t' \\
\state{C}{t}{\sigma}(x \altc y) =  
\state{C}{t}{\sigma}(x) \altc \state{C}{t}{\sigma}(y)              \\
\state{C}{t}{\sigma}(x \tout y) = 
\state{C}{t}{\sigma}(x) \tout \state{C}{t}{\sigma}(y) 
\eqnsep
\multicolumn{2}{@{}l@{}}
{\hspace*{4.5em} 
 \mbox{side-condition of all equation schemas in which $c$\, and $C$\,
       occur: $c \in C$,}} 
\\
\multicolumn{2}{@{}l@{}}
{\hspace*{4.5em} 
 \mbox{side-condition of all equation schemas in which $t'$ and $t''$
       occur: $t' < t''$}} 
\end{ceqncol}
\end{eqntbl}
\end{table}
In these tables, 
$\alpha$ stands for an arbitrary atomic process term from $\AProc$,
$a$ and $a'$ stand for arbitrary actions from $\ATAct \union \RTAct$,
$c$ stands for an arbitrary channel from $\Chan$,
$d$ stands for an arbitrary datum from $\Data$,
$t$, $t'$, and $t'''$ stand for arbitrary elements of $\Time$,
$t''$ stands for an arbitrary element of $\Time \union \set{\infty}$,
$\xi$ stands for an arbitrary element of $\Place$, 
$\sigma$ and $\sigma'$ stand for arbitrary communication states from 
$\State$, and
$C$ stands for an arbitrary subset of $\Chan$.
So, many equations in these tables are actually axiom schemas.
Side conditions restrict what $a$, $a'$, $c$, $d$, $t$, $t'$, $t''$, 
$t'''$, $\xi$, $\sigma$, $\sigma'$, and $C$ stand for. 

Most of the equations in Table~\ref{eqns-STPA-I} are reminiscent of 
axioms of \ACP$\rho\sigma$, the extension of \ACP\ to timed behaviour in 
space introduced in~\cite{BB93a}.
The first seven equations in Table~\ref{eqns-STPA-I} are the axioms of 
\BPAd, a subtheory of \ACP\ that does not cover parallelism 
and communication (see e.g.~\cite{BW90}).

Table~\ref{eqns-STPA-II} concerns the axioms for the state operators of 
\STPA.
These axioms are reminiscent of the axioms for the state operators used 
in~\cite{BB93a} to model, together with the integration operator, 
asynchronous communication in space-time.
However, in~\cite{BB93a}, the number of axiom schemas is kept small by 
introducing, as is customary with the axiomatization of state 
operators, so-called action and effect functions. 
Because it complicates determining whether an equation is an axiom, we 
are breaking with this custom here.
The differences with the axioms from~\cite{BB93a} are mainly caused by 
the different kind of potential receive action in \STPA, a kind that 
does not restrict the actual point of time at which a datum can be 
received to a single point in time.
Moreover, the state operators of \STPA\ combine the usual role of state
operators with the role of the initialization operator 
from~\cite{BB93a}.

The time-out operator cannot always be eliminated from process terms by 
means of the axioms of \STPA.
For example, the time-out operator cannot be eliminated from: 
\begin{ldispl}
\begin{tabular}[t]{@{}l@{\,\,}c@{\,\,}l@{}}
$\aprcv{c}{d}{t}{t'}{\xi} \tout \adead{t''}$ & if & 
$t \leq t'' < t'$, \\
$a \tout \aprcv{c}{d}{t}{t'}{\xi}$ & if & 
$\ubt(a) > t$ or $a \notin \ATAct$.  
\end{tabular}
\end{ldispl}%
In the first term, this is due to the presence of a potential receipt.
In the second term, this is due to the presence of a potential receipt 
or the presence of both absolute timing and relative timing.
The problem is that the point in time at which a potential receipt takes 
place is not fixed and that the initialization time needed to relate the 
two kinds of timing is not fixed.
This informally explains why the time-out operator cannot be eliminated 
from the above terms.
The question remains to what extent the time-out operator can be 
eliminated from process terms.
We will return to this question in Section~\ref{sect-sound-complete}.

Let $t,t' \in \Time$ be such that $t \leq t'$.
Then we can easily derive the following equation from the axioms of 
\STPA:
\begin{ldispl}
\state{\set{c}}{0}{\emptyset}
 (\apsnd{c}{d}{t}{\xi} \parc \aprcv{c}{d}{0}{t'}{\xi}) =
\aesnd{c}{d}{t}{\xi} \seqc \aercv{c}{d}{t}{\xi}\;.
\end{ldispl}%
This equation shows that if a process at some point in space sends a 
datum and another process at the same point in space receives that 
datum, the sending and receiving take place in that order, but at the 
same point in time.

Let $t,t' \in \Time$ be such that $t \leq t'$ and 
let $d_1,\ldots,d_n \in \Data$ be such that $d_1,\ldots,d_n$ are 
mutually different.
Then we can derive the following equation from the axioms of \STPA\
($1 \leq i \leq n$):
\begin{ldispl}
\state{\set{c}}{0}{\emptyset}
 (\apsnd{c}{d_i}{t}{\xi} \parc
  (\aprcv{c}{d_1}{0}{t'}{\xi} \altc \ldots \altc
   \aprcv{c}{d_n}{0}{t'}{\xi})) 
\\ 
\quad {} =
\aesnd{c}{d_i}{t}{\xi} \seqc (\aercv{c}{d_i}{t}{\xi} \altc \adead{t'})\;.
\end{ldispl}%
This equation shows that a process waiting to receive a datum may let 
pass the point in time that the datum arrives.
In other words, reception of a datum is not given priority over idling.
We will return to this issue in Section~\ref{sect-maxpr}.

The commutativity and associativity of the operator $\altc$ permit the 
use of the notation $\Altc{i \in I} T_i$, where 
$I = \set{i_1,\ldots,i_n}$, for the term 
$T_{i_1} \altc \ldots \altc T_{i_n}$.
The convention is used that $\Altc{i \in I} T_i$ stands for 
$\dead$ if $I = \emptyset$.
Moreover, we write $\Altc{i < n} T_i$, where $n \in \Nat$, for 
$\Altc{i \in \set{j \in \Nat \where j < n}} T_i$.

\section{Space-Time Process Algebra with Maximal Progress}
\label{sect-maxpr}

In the case of asynchronous communication in space-time, reception of a
datum is usually given priority over idling.
This calls for special priority operators, known as the maximal
progress operators (cf.~\cite{BB93a,BM03b}).
In this section, we extend \STPA\ by adding the special priority 
operators in question and axioms concerning these additional operators.
We write \STPAt\ for the resulting theory.

In this section will sometimes be referred to the subsets of $\AAct$ 
that consist of all actual actions timed at a certain point in time.

Let $t \in \Time$.
Then the set $\AAct(t)$ of actual actions timed at point in time 
$t$ is defined as follows:
\begin{ldispl}
\begin{aeqns}
\AAct(t) & = & \set{a \in \AAct \where \bt(a) = t}\;.
\end{aeqns} 
\end{ldispl}% 

In the case of \STPAt, priorities are given by a partial ordering 
$\prio{H}$ on $\AProc$ determined by a set $H \subseteq \AAct$.%
\footnote
{Recall that $\AProc$ is the set of atomic process terms and that
 the union of the set $\PAct$ of potential actions and the set $\AAct$ 
 of actual actions is a proper subset of $\AProc$.}
Informally, $\alpha \prio{H} \alpha'$ iff $\alpha$ is an actual action 
or an absolutely timed inaction, $\alpha'$ belongs to $H$, and either 
$\alpha$ idles longer than $\alpha'$ or $\alpha$ idles as long as 
$\alpha'$ and does not belong to $H$.

Let $H \subseteq \AAct$.
Then, for all $\alpha,\alpha' \in \AProc$, $\alpha \prio{H} \alpha'$ iff 
one of the following conditions holds:
\begin{itemize}
\item
there exist $t,t' \in \Time$ with $t < t'$ such that 
$\alpha \in \AAct(t') \union \set{\adead{t'}}$, $\alpha' \in \AAct(t)$, 
and $\alpha' \in H$;
\item
there exists a $t \in \Time$ such that $\alpha,\alpha' \in \AAct(t)$, 
$\alpha \notin H$, and $\alpha' \in H$.
\end{itemize}

The signature of \STPAt\ is the signature of \STPA\ with, for each 
$H \subseteq \AAct$, added: 
\begin{itemize}
\item
the unary \emph{maximal progress} operator $\maxpr{H}$;
\item
the binary \emph{maximal progress} operator $\auxpr{H}$.
\end{itemize}

Let $H \subseteq \AAct$ and $P \in \Proc$.
Intuitively, the maximal progress operator $\maxpr{H}(P)$ can be 
explained as follows:
\begin{itemize}
\item
$\maxpr{H}(P)$ behaves the same as $P$ except that performing an action 
from $H$ has priority over idling and over performing an action not from 
$H$ whenever such alternatives occur.
\end{itemize}
For each $H \subseteq \AAct$, the operator $\auxpr{H}$ is a convenient 
auxiliary operator for the axiomatization of $\maxpr{H}$.
The operator $\auxpr{H}$ is inspired by the operator $\triangle$ used 
in~\cite{ABV94a} to axiomatize a priority operator in an untimed 
setting.
In $\maxpr{H}(P)$, $P$ has two roles: it provides both the high-priority 
behaviour of $P$ and the low-priority behaviour of $P$ that is blocked 
by the high-priority behaviour of $P$. 
The binary priority operator $\auxpr{H}$ separates the two roles of $P$ 
in $\maxpr{H}(P)$: in $P \auxpr{H} Q$, the low-priority behaviour of $P$ 
is blocked by the high-priority behaviour of $Q$.

The axiom system of \STPAt\ is the axiom system of \STPA\ with the 
axioms for the maximal progress operators added.
These additional axioms are given in Table~\ref{eqns-maxpr}.%
\begin{table}[!t]
\caption{Additional axioms for the maximal progress operators}
\label{eqns-maxpr}
\begin{eqntbl}
\begin{ceqncol}
\maxpr{H}(x)  = x \auxpr{H} x 
\eqnsep
\alpha \auxpr{H} \alpha' = \alpha & \mif \alpha \nprio{H} \alpha' \\
\alpha \auxpr{H} \alpha' = \adead{t} 
        & \mif \alpha \prio{H}  \alpha' \And \alpha' \in \AAct(t) \\
x \seqc y \auxpr{H} z   = (x \auxpr{H} z) \seqc \maxpr{H}(y)      \\
(x \altc y) \auxpr{H} z = (x \auxpr{H} z) \altc (y \auxpr{H} z)   \\
x \auxpr{H} y \seqc z   = x \auxpr{H} y                           \\
x \auxpr{H} (y \altc z) = (x \auxpr{H} y) \auxpr{H} z 
\end{ceqncol}
\end{eqntbl}
\end{table}
In this table, 
$H$ stands for an arbitrary subset of $\AAct$,
$\alpha$ and $\alpha'$ stand for arbitrary atomic process terms from 
$\AProc$, and 
$t$ stands for an arbitrary element of $\Time$.

Let $c \in \Chan$, $d \in \Data$, $\xi \in \Place$, 
$H = \set{\aercv{c}{d}{t}{\xi} \where t \in \Time}$, and
$t,t' \in \Time$ be such that $t < t'$.
Then we can easily derive the following equation from the axioms of 
\STPAt:
\begin{ldispl}
\maxpr{H}(\aercv{c}{d}{t}{\xi} \altc \aercv{c}{d}{t'}{\xi} \altc
          \aesnd{c}{d}{t}{\xi}) = \aercv{c}{d}{t}{\xi}\;.
\end{ldispl}%
This corresponds to the intuition about the priority ordering 
$\prio{H}$: $\aercv{c}{d}{t'}{\xi}$ idles longer than 
$\aercv{c}{d}{t}{\xi}$ and $\aesnd{c}{d}{t}{\xi}$ does not belong to 
$H$.

\section{Space-Time Process Algebra with Guarded Recursion}
\label{sect-recursion}

A closed term over the signature of \STPA\ denotes a process with a 
finite upper bound to the number of actions that it can perform. 
Guarded recursion allows the description of processes without a finite 
upper bound to the number of actions that it can perform.
In this section, we extend \STPA\ with guarded recursion by adding 
constants for solutions of guarded recursive specifications and axioms 
concerning these additional constants.
We write \STPA\REC\ for the resulting theory.

Let $X$ be a variable from $\cX$, and 
let $T$ be a term over the signature of \STPA\ in which $X$ occurs.
Then an occurrence of $X$ in $T$ is \emph{guarded} if $T$ has a subterm 
of the form $a \seqc T'$ where $a \in \PAct \union \AAct$ and $T'$ 
contains this occurrence of $X$.
A term $T$ over the signature of \STPA\ is \emph{guarded} if all 
occurrences of variables in $T$ are guarded. 

A \emph{recursive specification} over the signature of \STPA\ is a set 
$\set{X_i = T_i \where i \in I}$, 
where $I$ is an index set, 
each $X_i$ is a variable from $\cX$, 
each $T_i$ is a term over the signature of \STPA\ in which only 
variables from $\set{X_i \where i \in I}$ occur, 
and $X_i \neq X_j$ for all $i,j \in I$ with $i \neq j$.
We write $\vars(E)$, where $E$ is a recursive specification
$\set{X_i = T_i \where i \in I}$ over the signature of \STPA, for the 
set $\set{X_i \where i \in I}$.
\begin{comment}
%
Let $E$ be a recursive specification over the signature of \STPA\ and 
let $X \in \vars(E)$. 
Then the unique equation $X = T\; \in \;E$ is called the 
\emph{recursion equation for $X$ in $E$}.
\end{comment}

A recursive specification $\set{X_i = T_i \where i \in I}$ over the 
signature of \STPA\ is \emph{guarded} if each $T_i$ is rewritable to a 
guarded term by using the axioms of \STPA\ in either direction and/or 
the equations in $\set{X_j = T_j \where j \in I \And i \neq j}$ from 
left to right.

Let $\set{X_i = T_i \where i \in I}$ be a recursive specification over 
the signature of \STPA.
Then a solution of $\set{X_i = T_i \where i \in I}$ in some model of 
\STPA\ is a set $\set{p_i \where i \in I}$ of processes 
in the model such that each equation in 
$\set{X_i = T_i \where i \in I}$ holds in the model if, for each 
$i \in I$, $X_i$ is assigned $p_i$.
Here, $p_i$ is said to be the \emph{$X_i$-component} of the solution.
A guarded recursive specification has a unique solution in the intended 
model of \STPA, to wit the bisimulation model of \STPA\ defined in 
Section~\ref{sect-SOS}.
A recursive specification that is not guarded may not have a unique 
solution in that model.

Below, for each recursive specification $E$ over the signature of \STPA\ 
that is guarded and $X \in \vars(E)$, a constant $\rec{X}{E}$ that 
stands for the $X$-component of the unique solution of $E$ will be 
introduced. 
The notation $\rec{T}{E}$ will be used for $T$ with, for all 
$X \in \vars(E)$, all occurrences of $X$ in $T$ replaced by 
$\rec{X}{E}$.

The signature of \STPA\REC\ is the signature of \STPA\ with, for each 
guarded recursive specification $E$ over the signature of \STPA\ and 
$X \in \vars(E)$, added a \emph{recursion} constant $\rec{X}{E}$.

The axiom system of \STPA\REC\ is the axiom system of \STPA\ with, for 
each guarded recursive specification $E$ over the signature of \STPA\ 
and $X \in \vars(E)$, added the equation $\rec{X}{E} = \rec{T}{E}$ for 
the unique term $T$ over the signature of \STPA\ such that 
$X = T \,\in\, E$ and the conditional equation 
$E \,\Implies\, X = \rec{X}{E}$.

The equations and conditional equations added to the axiom system of 
\STPA\ to obtain the axiom system of \STPA\REC\ are the instances of the 
axiom schemas RDP and RSP, respectively, given in 
Table~\ref{axioms-REC}.
\begin{table}[!t]
\caption{Additional axioms for the recursion constants}
\label{axioms-REC}
\begin{eqntbl}
\begin{caxcol}
\rec{X}{E} = \rec{T}{E}          & \mif X = T \,\in\, E & \axiom{RDP} \\
E \,\Implies\, X = \rec{X}{E}    & \mif X \in \vars(E)  & \axiom{RSP} 
\end{caxcol}
\end{eqntbl}
\end{table}
In these axiom schemas, 
$X$ stands for an arbitrary variable from $\cX$, 
$T$ stands for an arbitrary term over the signature of \STPA, and
$E$ stands for an arbitrary guarded recursive specification over the 
signature of \STPA.
Side conditions restrict what $X$, $T$ and $E$ stand for.

We write $\Procr$ for the set of all closed terms over the signature of 
\STPA\REC.

About RDP and RSP we remark that, for a fixed $E$, the equations 
$\rec{X}{E} = \rec{T}{E}$ and the conditional equations 
$E \Implies X \!=\! \rec{X}{E}$ express that the constants $\rec{X}{E}$ 
make up a solution of $E$ and that this solution is the only one.

Because conditional equations must be dealt with in \STPA\REC, it is 
understood that conditional equational logic is used in deriving 
equations from the axioms of \STPA\REC.
A complete inference system for conditional equational logic can be 
found in~\cite{BW90,Gog21a}.

We write $\mathit{Th} \Ent T = T'$, where $\mathit{Th}$ is \STPA\REC\ or 
\STPAt\REC, to indicate that the equation $T = T'$ is derivable from the 
axioms of $\mathit{Th}$ using a complete inference system for 
conditional equational logic.

We often write $X$ for $\rec{X}{E}$ if $E$ is clear from the context. 
In such cases, it should also be clear from the context that we use $X$ 
as a constant.

A special kind of guarded recursive specifications are linear recursive
specifications.
The right-hand sides of the equations in a linear recursive 
specification are terms of a special form.
The set $\LProc$ of \emph{linear} terms over the signature of \STPA\ is 
the smallest set satisfying the following rules:
\begin{itemize}
\item
if $t \in \Time$, then $\adead{t} \in \LProc$;
\item
if $a \in \AAct$, then $a \in \LProc$;
\item
if $a \in \AAct$ and $X \in \cX$, then $a \seqc X \in \LProc$;
\item
if $T,T' \in \LProc$, then $T \altc T' \in \LProc$.
\end{itemize}
A recursive specification $\set{X_i = T_i \where i \in I}$ over the 
signature of \STPA\ is \emph{linear} if each $T_i$ is a linear term over 
the signature of \STPA.
Obviously, all linear recursive specifications are guarded.
For recursion constants $\rec{X}{E}$ where $E$ is linear, the operational 
semantics of $\rec{X}{E}$ given in Section~\ref{sect-SOS} is well
reflected by $E$.
This is used in Section~\ref{sect-sound-complete} in the proof of a 
(semi-)completeness result.

\STPAt\ can be extended with guarded recursion in the same way as \STPA.
We write \STPAt\REC\ for the resulting theory.

\section{An Example: A Data Communication Protocol}
\label{sect-example}

\STPAt\REC\ is used in this section to describe an asynchronous version 
of the data communication protocol known as the \emph{PAR} (Positive 
Acknowledgement with Retransmission) protocol.

The configuration of the PAR protocol is shown in Fig.~\ref{fig-par}
by means of a connection diagram.
\begin{figure}
\centering
\setlength{\unitlength}{.18em}
\begin{picture}(150,47)(0,0)
\put(15,20){\line(1,0){10}}
\put(17,24){\makebox(0,0){$\ch_1$}}
\put(35,20){\circle{20}}
\put(35,20){\makebox(0,0){$S$}}
\put(43,25){\line(2,1){22}}
\put(52,34){\makebox(0,0){$\ch_3$}}
\put(75,36){\oval(20,10)}
\put(75,36){\makebox(0,0){$K$}}
\put(85,36){\line(2,-1){22}}
\put(98,34){\makebox(0,0){$\ch_4$}}
\put(43,15){\line(2,-1){22}}
\put(52,6){\makebox(0,0){$\ch_5$}}
\put(75,4){\oval(20,10)}
\put(75,4){\makebox(0,0){$L$}}
\put(85,4){\line(2,1){22}}
\put(98,6){\makebox(0,0){$\ch_6$}}
\put(115,20){\circle{20}}
\put(115,20){\makebox(0,0){$R$}}
\put(125,20){\line(1,0){10}}
\put(133,24){\makebox(0,0){$\ch_2$}}
\end{picture}
\caption{Connection diagram for the PAR protocol}
\label{fig-par}
\end{figure}
The sender waits for an acknowledgement before a new datum is
transmitted.
If an acknowledgement is not received within a complete protocol cycle,
the old datum is retransmitted.
In order to avoid duplicates due to retransmission, data are labeled
with an alternating bit from $B = \set{0,1}$.

We have a sender process $S$, a receiver process $R$, and two repeater
processes $K$ and $L$.
Process $S$ waits until a datum $d$ is offered on external 
channel~$\ch_1$.
When a datum is offered on this channel, $S$ consumes it, packs it with 
an alternating bit $b$ in a frame $\tup{d,b}$, and then delivers the 
frame on channel~$\ch_3$.
Next, $S$ waits until an acknowledgement $\ack$ is offered on 
channel~$\ch_5$.
When the acknowledgement does not arrive within a certain time period,
$S$ delivers the same frame again and goes back to waiting for an
acknowledgement.
When the acknowledgement arrives within that time period, $S$ goes back
to waiting for a datum.
Process $R$ waits until a frame with a datum and an alternating bit
$\tup{d,b}$ is offered on channel~$\ch_4$.
When a frame is offered on this channel, $R$ consumes it, unpacks it, 
and then delivers the datum $d$ on channel~$\ch_2$ if the alternating bit 
$b$ is the right one and in any case an acknowledgement $\ack$ at 
channel~$\ch_6$.
After that, $R$ goes back to waiting for a frame, but the right bit
changes to $(1-b)$ if the alternating bit was the right one.
Processes $K$ and $L$ pass on frames from channel $\ch_3$ to channel 
$\ch_4$ and acknowledgements from channel $\ch_6$ to channel $\ch_5$, 
respectively.
The repeaters may produce an error instead of passing on frames or 
acknowledgements.
The times $t_S$, $t_R$, $t_K$, and $t_L$ are the times that it takes the 
different processes to pack and deliver, to unpack and deliver or simply 
to deliver what they consume.
The time $t_S'$ is the time-out time of the sender, i.e., the time
after which it retransmits a datum in case it is still waiting for an
acknowledgement.
The time $t_R'$ is the time that it takes the receiver to produce and
deliver an acknowledgement.
The points in space $\xi_S$, $\xi_R$, $\xi_K$, and $\xi_L$ are the 
points in space at which the different processes take place.

We assume that a finite set $D$ of data such that $D \subset \Data$ and
$D \x B \subset \Data$ has been given.
Moreover, we assume that $\ack \in \Data$ and $\err \in \Data$.

Below, we give the recursive specifications of $S$, $R$, $K$, and $L$.
We refrain from mentioning after each equation schema that there is an
instance for every $d \in D$ and/or $b \in B$.

% \noindent
The recursive specification of the sender $S$ consists of the following
equations:
\begin{ldispl}
\begin{aeqns}
S         & = & S_0\;,
\\
S_b       & = &
\Altc{d \in D} \rprcv{\ch_1}{d}{0}{\infty}{\xi_S} \seqc S'_{d,b}\;, 
\\
S'_{d,b}  & = & 
\rpsnd{\ch_3}{(d,b)}{t_S}{\xi_S} \seqc S''_{d,b}\;,
\\
S''_{d,b} & = &
\rprcv{\ch_5}{\ack}{0}{t'_S}{\xi_S} \seqc S_{1{-}b} \altc
\rpsnd{\ch_3}{(d,b)}{t'_S}{\xi_S} \seqc S''_{d,b}\;, 
\end{aeqns}
\end{ldispl}%
the recursive specification of the receiver $R$ consists of the 
following equations:
\begin{ldispl}
\begin{aeqns}
R         & = & R_0\;,
\\
R_b       & = &
\Altc{d \in D} \rprcv{\ch_4}{(d,b)}{0}{\infty}{\xi_R} \seqc R'_{d,b}
 \altc
\Altc{d \in D} \rprcv{\ch_4}{(d,1-b)}{0}{\infty}{\xi_R} \seqc R''_b
 \;,
\\
R'_{d,b}  & = & 
\rpsnd{\ch_2}{d}{t_R}{\xi_R} \seqc R''_{1{-}b}\;, 
\\
R''_b     & = &
\rpsnd{\ch_6}{\ack}{t'_R}{\xi_R} \seqc R_b\;,
\end{aeqns}
\end{ldispl}%
the recursive specification of the repeater $K$ consists of the 
following equations:
\begin{ldispl}
\begin{aeqns}
K         & = & 
\Altc{(d,b) \in D \x B}
 \rprcv{\ch_3}{(d,b)}{0}{\infty}{\xi_K} \seqc K'_{d,b}\;,
\\
K'_{d,b}  & = &
\rpsnd{\ch_4}{(d,b)}{t_K}{\xi_K} \seqc K \altc 
\rpsnd{\ch_4}{\err}{t_K}{\xi_K} \seqc K\;, 
\end{aeqns}
\end{ldispl}%
and the recursive specification of the repeater $L$ consists of the 
following equations:
\begin{ldispl}
\begin{aeqns}
L         & = & 
\rprcv{\ch_6}{\ack}{0}{\infty}{\xi_L} \seqc L'\;,
\\
L'        & = &
\rpsnd{\ch_5}{\ack}{t_L}{\xi_L} \seqc L \altc 
\rpsnd{\ch_5}{\err}{t_L}{\xi_L} \seqc L\;.
\end{aeqns}
\end{ldispl}%
The whole protocol is described by the term
\begin{ldispl}
\maxpr{H}
 (\state{\set{\ch_3,\ch_4,\ch_5,\ch_6}}{0}{\emptyset}
   (S \parc K \parc L \parc R))\;,
\end{ldispl}%
where
$H =
 \set{\aercv{c}{d}{t}{\xi} \where
      c \in \set{\ch_3,\ch_4,\ch_5,\ch_6} \And d \in \Data \And
      t \in \Time \And \xi \in \Place}$.

In the system described by the term
\begin{ldispl}
\state{\set{\ch_3,\ch_4,\ch_5,\ch_6}}{0}{\emptyset}
 (S \parc K \parc L \parc R)\;,
\end{ldispl}% 
when $R$ is ready to receive a datum that $S$ has sent at point in space 
$\xi_S$, $R$ may let pass the point in time that the datum arrives at 
point in space $\xi_R$ because reception of a datum is not given
priority over idling.
By using a maximal progress operator, this anomaly is not present in the 
protocol as described earlier.

A necessary condition for this protocol to be correct is that the 
time-out time $t_S'$ is longer than a complete protocol cycle, i.e. 
\begin{ldispl}
t_S' > 
\frac{\dist(\xi_S,\xi_K)}{v} + t_K + \frac{\dist(\xi_K,\xi_R)}{v} +
t_R + t_R' + 
\frac{\dist(\xi_R,\xi_L)}{v} + t_L + \frac{\dist(\xi_L,\xi_S)}{v}\;.
\end{ldispl}% 
If the time-out time is shorter than a complete protocol cycle, the
time-out is called premature.
In that case, while an acknowledgement is still on the way, the sender
will retransmit the current frame.
When the acknowledgement finally arrives, the sender will treat this
acknowledgement as an acknowledgement of the retransmitted frame.
However, an acknowledgement of the retransmitted frame may be on the
way.
If the next frame transmitted gets lost and the latter acknowledgement
arrives, no retransmission of that frame will follow and the protocol
will fail.

In this paper, the focus is on asynchronous communication in space-time.
However, \STPA\ can be easily extended with the spatial replacement 
operators from~\cite{BB93a,BM03b} to deal with processes that move in 
space.
This would allow us to describe a variant of the protocol described 
above where the repeater processes $K$ and $L$ move in space.
The state operators of \STPA\ can also be easily adapted to deal 
uniformly with all transmission limitations caused by blocking solid 
objects (as in~\cite{BM03b}). 

In~\cite{BM02a}, a synchronous version of the PAR protocol is described
and analyzed using a generalization of \ACP\ in which time is measured 
on a discrete time scale and spatial distribution is ignored.%
\footnote
{The treatment of that version of the PAR protocol has been copied
 almost verbatim without mentioning its origin in at least one other
 publication.}
The treatment of an asynchronous version of the PAR protocol in this 
section is based on the treatment of that synchronous version 
in~\cite{BM02a}.
In the case of the synchronous version of the PAR protocol 
from~\cite{BM02a}, the necessary condition for correctness becomes
\begin{ldispl}
t_S' > t_K + t_R + t_R' + t_L\;,
\end{ldispl}% 
which seems weaker than the one mentioned earlier. 

One view of the synchronous version of the PAR protocol is that it is 
odd: synchronous communication is possible only if there is no spatial
distribution, but the protocol is useless without a spatial 
distribution. 
Another view is the following: the communication is in fact 
asynchronous, the process $K$ is an abstraction of everything that takes 
place between sender and receiver to pass on frames and the process $L$ 
is an abstraction of everything that takes place between receiver and 
sender to pass on acknowledgements.

Under the latter view, $t_K$ must include the transmission times to and 
from $K$ and $t_L$ must include the transmission times to and from $L$ 
--- in which case the above necessary condition for correctness is not 
really weaker than the one mentioned earlier.
However, an issue with applying this view is that the protocol 
description in~\cite{BM02a} does not contain any details that indicate 
that asynchronous communication in space-time is involved.
Much detail has to be added to the description, which is not possible
in the process algebra used anyway, before it can be adapted to the case 
where, for example, $K$ and/or $L$ move in space.

\section{Operational Semantics and Bisimilarity}
\label{sect-SOS}

In this section, we give a structural operational semantics for \STPA, 
and we define a notion of bisimilarity based on the structural 
operational semantics.

The structural operational semantics for \STPA\ consists of the
following transition relations:
\begin{itemize}
\item
a binary relation ${\step{\ell}}$ on $\Procr$ 
for each $\ell \in \Time \x \State \x \AAct$;
\item
a unary relation ${\step{\ell}\!\term}$ on $\Procr$ 
for each $\ell \in \Time \x \State \x \AAct$;
\item
a unary relation ${\idle{\ell}}$ on $\Procr$ 
for each $\ell \in \Time \x \State \x \Time$.
\end{itemize}
We write
$\astp{P}{t,\sigma}{a}{Q}$ for $(P,Q) \in {\step{(t,\sigma,a)}}$,
$\astp{P}{t,\sigma}{a}{\!\term}$ for $P \in {\step{(t,\sigma,a)}\!\term}$, and
$\istp{P}{t,\sigma}{t'}$ for $P \in {\idle{(t,\sigma,t')}}$.

Let $P,Q \in \Procr$, $c \in \Chan$, $d \in \Data$, $t,t' \in \Time$, and
$\sigma \in \State$.
Then the transition relations introduced above can be explained as 
follows:
\begin{itemize}
\item
$\astp{P}{t,\sigma}{\aesnd{c}{d}{t'}{\xi}}{Q}$:\,
if the point in time is $t$ and the communication state is~$\sigma$, 
then the process denoted by $P$ is capable of making a transition 
to the process denoted by $Q$ by sending datum $d$ at point in time $t'$ 
and point in space $\xi$;
\item
$\astp{P}{t,\sigma}{\aercv{c}{d}{t'}{\xi}}{Q}$:\,
if the point in time is $t$ and the communication state is~$\sigma$, 
then the process denoted by $P$ is capable of making a transition 
to \linebreak[2] the \linebreak[2] process denoted by $Q$ by receiving 
datum $d$ at point in time $t'$ and point in space $\xi$;
\item
$\astp{P}{t,\sigma}{\aesnd{c}{d}{t'}{\xi}}{\!\term}$:\,
if the point in time is $t$ and the communication state is~$\sigma$, 
then the process denoted by $P$ is capable of terminating 
successfully after sending datum $d$ at point in time $t'$ and point in 
space $\xi$;
\item
$\astp{P}{t,\sigma}{\aercv{c}{d}{t'}{\xi}}{\!\term}$:\,
if the point in time is $t$ and the communication state is~$\sigma$, 
then the process denoted by $P$ is capable of terminating 
successfully after receiving datum $d$ at point in time $t'$ and point 
in space $\xi$;
\item
$\istp{P}{t,\sigma}{t'}$:\,
if the point in time is $t$ and the communication state is $\sigma$, 
then the process denoted by $P$ is capable of idling till point in 
time $t'$.
\end{itemize}

The structural operational semantics of \STPA\ is described by the rules
given in Tables~\ref{rules-STPA-I} and~\ref{rules-STPA-II}.%
\begin{table}[!b]
\caption{Operational semantics for \STPA\ (Part I)}
\label{rules-STPA-I}
\begin{ruletbl}
\Rule
{[t \leq t'' \leq t']}
{\istp{\adead{t'}}{t,\sigma}{t''}}
\quad
\Rule
{[t'' \leq t']}
{\istp{\rdead{t'}}{t,\sigma}{t + t''}}
\\
\Rule
{[t \leq t']}
{\astp{\apsnd{c}{d}{t'}{\xi}}
  {t,\sigma}{\aesnd{c}{d}{t'}{\xi}}{\!\term}}
\quad
\Rule
{[t \leq t'' \leq t']}
{\istp{\apsnd{c}{d}{t'}{\xi}}{t,\sigma}{t''}}
\\
\Rule
{\phantom{[t \leq t']}}
{\astp{\rpsnd{c}{d}{t'}{\xi}}
  {t,\sigma}{\aesnd{c}{d}{t + t'}{\xi}}{\!\term}}
\quad
\Rule
{[t'' \leq t']}
{\istp{\rpsnd{c}{d}{t'}{\xi}}{t,\sigma}{t + t''}}
\\
\Rule
{[t < t'',\; t' < t'',\; V = \rcpt(\sigma,c,d,\max(t,t'),t'',\xi),\;
  V \neq \emptyset]}
{\astp{\aprcv{c}{d}{t'}{t''}{\xi}}
  {t,\sigma}{\aercv{c}{d}{\min(V)}{\xi}}{\!\term}}
\\
\Rule
{[t < t'',\; t' < t'',\; V = \rcpt(\sigma,c,d,\max(t,t'),t'',\xi),\;
  V \neq \emptyset,\; t''' \leq \min(V)]}
{\istp{\aprcv{c}{d}{t'}{t''}{\xi}}{t,\sigma}{t'''}}
\\
\Rule
{[t < t'',\; t' < t'',\; V = \rcpt(\sigma,c,d,\max(t,t'),t'',\xi),\;
  V = \emptyset,\; t \leq t''' \leq t'']}
{\istp{\aprcv{c}{d}{t'}{t''}{\xi}}{t,\sigma}{t'''}}
\\
\Rule
{[t' < t'',\; V = \rcpt(\sigma,c,d,t+t',t+t'',\xi),\; 
  V \neq \emptyset]}
{\astp{\rprcv{c}{d}{t'}{t''}{\xi}}
  {t,\sigma}{\aercv{c}{d}{\min(V)}{\xi}}{\!\term}}
\\
\Rule
{[t' < t'',\; V = \rcpt(\sigma,c,d,t+t',t+t'',\xi),\; 
  V \neq \emptyset,\; t''' \leq \min(V)]}
{\istp{\rprcv{c}{d}{t'}{t''}{\xi}}{t,\sigma}{t'''}}
\\
\Rule
{[t' < t'',\; V = \rcpt(\sigma,c,d,t+t',t+t'',\xi),\;
  V = \emptyset,\; t' \leq t''' \leq t'']}
{\istp{\rprcv{c}{d}{t'}{t''}{\xi}}{t,\sigma}{t+t'''}}
\\
\Rule
{[t \leq t']}
{\astp{\aesnd{c}{d}{t'}{\xi}}{t,\sigma}{\aesnd{c}{d}{t'}{\xi}}{\!\term}}
\quad
\Rule
{[t \leq t'' \leq t']}
{\istp{\aesnd{c}{d}{t'}{\xi}}{t,\sigma}{t''}}
\\
\Rule
{[t \leq t']}
{\astp{\aercv{c}{d}{t'}{\xi}}{t,\sigma}{\aercv{c}{d}{t'}{\xi}}{\!\term}}
\quad
\Rule
{[t \leq t'' \leq t']}
{\istp{\aercv{c}{d}{t'}{\xi}}{t,\sigma}{t''}}
\end{ruletbl}
\end{table}
\begin{table}[!p]
\caption{Operational semantics for \STPA\ (Part II)}
\label{rules-STPA-II}
\begin{ruletbl}
\Rule
{\astp{x}{t,\sigma}{a}{x'}}
{\astp{x \altc y}{t,\sigma}{a}{x'}}
\quad
\Rule
{\astp{x}{t,\sigma}{a}{\!\term}}
{\astp{x \altc y}{t,\sigma}{a}{\!\term}}
\quad
\Rule
{\astp{y}{t,\sigma}{a}{y'}}
{\astp{x \altc y}{t,\sigma}{a}{y'}}
\quad
\Rule
{\astp{y}{t,\sigma}{a}{\!\term}}
{\astp{x \altc y}{t,\sigma}{a}{\!\term}}
\\
\Rule
{\istp{x}{t,\sigma}{t'}}
{\istp{x \altc y}{t,\sigma}{t'}}
\quad
\Rule
{\istp{y}{t,\sigma}{t'}}
{\istp{x \altc y}{t,\sigma}{t'}}
\\
\Rule
{\astp{x}{t,\sigma}{a}{x'}}
{\astp{x \seqc y}{t,\sigma}{a}{x' \seqc y}}
\quad
\Rule
{\astp{x}{t,\sigma}{a}{\!\term}}
{\astp{x \seqc y}{t,\sigma}{a}{y}}
\quad
\Rule
{\istp{x}{t,\sigma}{t'}}
{\istp{x \seqc y}{t,\sigma}{t'}}
\\
\Rule
{\astp{x}{t,\sigma}{a}{x'},\; \istp{y}{t,\sigma}{t'}\;
 [\bt(a) = t']}
{\astp{x \parc y}{t,\sigma}{a}{x' \parc y}}
\quad
\Rule
{\astp{x}{t,\sigma}{a}{\!\term},\; \istp{y}{t,\sigma}{t'}\;
 [\bt(a) = t']}
{\astp{x \parc y}{t,\sigma}{a}{y}}
\\
\Rule
{\istp{x}{t,\sigma}{t'},\,\; \astp{y}{t,\sigma}{a}{y'}\;
 [\bt(a) = t']}
{\astp{x \parc y}{t,\sigma}{a}{x \parc y'}}
\quad
\Rule
{\istp{x}{t,\sigma}{t'},\,\; \astp{y}{t,\sigma}{a}{\!\term}\,\;
 [\bt(a) = t']}
{\astp{x \parc y}{t,\sigma}{a}{x}}
\\
\Rule
{\istp{x}{t,\sigma}{t'},\; \istp{y}{t,\sigma}{t'}}
{\istp{x \parc y}{t,\sigma}{t'}}
\\
\Rule
{\astp{x}{t,\sigma}{a}{x'},\; \istp{y}{t,\sigma}{t'}\;
 [\bt(a) = t']}
{\astp{x \leftm y}{t,\sigma}{a}{x' \parc y}}
\quad
\Rule
{\astp{x}{t,\sigma}{a}{\!\term},\; \istp{y}{t,\sigma}{t'}\;
 [\bt(a) = t']}
{\astp{x \leftm y}{t,\sigma}{a}{y}}
\\
\Rule
{\istp{x}{t,\sigma}{t'},\; \istp{y}{t,\sigma}{t'}}
{\istp{x \leftm y}{t,\sigma}{t'}}
\\
\Rule
{\astp{x}{t,\sigma}{a}{x'},\; \istp{y}{t,\sigma}{t'}\;
 [\bt(a) = t']}
{\astp{x \tout y}{t,\sigma}{a}{x'}}
\quad
\Rule
{\astp{x}{t,\sigma}{a}{\!\term},\; \istp{y}{t,\sigma}{t'}\;
 [\bt(a) = t']}
{\astp{x \tout y}{t,\sigma}{a}{\!\term}}
\\
\Rule
{\istp{x}{t,\sigma}{t'},\; \istp{y}{t,\sigma}{t'}}
{\istp{x \tout y}{t,\sigma}{t'}}
\\
\Rule
{\astp{x}{t,\sigma}{a}{x'}\; [\chan(a) \notin C]}
{\astp{\state{C}{t}{\sigma}(x)}{t,\sigma}{a}{\state{C}{t}{\sigma}(x')}}
\quad
\Rule
{\astp{x}{t,\sigma}{a}{\!\term}\,\; [\chan(a) \notin C]}
{\astp{\state{C}{t}{\sigma}(x)}{t,\sigma}{a}{\!\term}}
\quad
\Rule
{\istp{x}{t,\sigma}{t'}}
{\istp{\state{C}{t}{\sigma}(x)}{t,\sigma}{t'}}
\\
\Rule
{\astp{x}{t,\sigma}{\aesnd{c}{d}{t'}{\xi}}{x'}\; [c \in C]}
{\astp{\state{C}{t}{\sigma}(x)}{t,\sigma}{\aesnd{c}{d}{t'}{\xi}}
      {\state{C}{t'}{\sigma \union \set{(c,d,t',\xi)}}(x')}}
\quad
\Rule
{\astp{x}{t,\sigma}{\aesnd{c}{d}{t'}{\xi}}{\!\term}\,\; [c \in C]}
{\astp{\state{C}{t}{\sigma}(x)}{t,\sigma}{\aesnd{c}{d}{t'}{\xi}}
      {\!\term}}
\\
\Rule
{\astp{x}{t,\sigma}{\aercv{c}{d}{t'}{\xi}}{x'}\; [c \in C]}
{\astp{\state{C}{t}{\sigma}(x)}{t,\sigma}{\aercv{c}{d}{t'}{\xi}}
      {\state{C}{t'}{\sigma}(x')}}
\quad
\Rule
{\astp{x}{t,\sigma}{\aercv{c}{d}{t'}{\xi}}{\!\term}\,\; [c \in C]}
{\astp{\state{C}{t}{\sigma}(x)}{t,\sigma}{\aercv{c}{d}{t'}{\xi}}
      {\!\term}}
\quad
\Rule
{[t' \leq t]}
{\istp{\state{C}{t}{\sigma}(x)}{t,\sigma}{t'}}
\end{ruletbl}
\end{table}
In these tables, 
$c$ stands for an arbitrary channel from $\Chan$, \linebreak[2]
$d$ stands for an arbitrary datum from $\Data$,\,
$t$, $t'$, $t''$, and $t'''$ stand for arbitrary elements of $\Time$,
$\xi$ stands for an arbitrary element of $\Place$, 
$V$ stands for an arbitrary subset of $\Time$, 
$a$ stands for an arbitrary actual action from $\AAct$,
$\sigma$ stands for an arbitrary communication state from $\State$, and
$C$ stands for an arbitrary subset of $\Chan$. 
So, many equations in these tables are actually rule schemas.
Side conditions restrict what $c$, $d$, $t$, $t'$, $t''$, $t'''$, $\xi$, 
$V$, $a$, $\sigma$, and $C$ stand for.

The rules in Tables~\ref{rules-STPA-I} and~\ref{rules-STPA-II} have the 
form \smash{\small $\SRule{\phi_1,\ldots,\phi_n\,[s]}{\phi}$}, where $[s]$ is optional.
They are to be read as ``if $\phi_1$ and \ldots and $\phi_n$ then 
$\phi$, provided $s$''.
As usual, $\phi_1,\ldots,\phi_n$ are called the premises and $\phi$ is 
called the conclusion.
A side condition $s$, if present, serves to restrict the applicability 
of a rule.
If a rule has no premises and no side-conditions, then nothing is 
displayed above the horizontal bar.

Let $\phi$ be \smash{\small $\astp{P}{t,\sigma}{a}{Q}$} or
\smash{\small $\astp{P}{t,\sigma}{a}{\!\term}$} or 
\smash{\small $\istp{P}{t,\sigma}{t'}$}.
Then, because the rules in Tables~\ref{rules-STPA-I} 
and~\ref{rules-STPA-II} constitute an inductive definition, 
$\phi$ holds iff it can be inferred from these rules.

Two processes are considered equal if they can simulate each other 
insofar as their capabilities to make transitions, to terminate 
successfully, and to idle are concerned.
This is covered by the notion of bisimilarity introduced below.

A \emph{bisimulation} is a symmetric relation 
$R \subseteq \Procr \x \Procr$ such that, 
for all $P, Q \in \Procr$ with $(P,Q) \in R$:
\begin{itemize}
\item
if $\astp{P}{t,\sigma}{a}{P'}$, then 
there exists a $Q' \in \Procr$ such that $\astp{Q}{t,\sigma}{a}{Q'}$ and 
$(P',Q') \in R$;
\item
if $\astp{P}{t,\sigma}{a}{\!\term}$, then 
$\astp{Q}{t,\sigma}{a}{\!\term}$; 
\item
if $\istp{P}{t,\sigma}{t'}$, then 
$\istp{Q}{t,\sigma}{t'}$.
\end{itemize}
Two closed terms $P, Q \in \Procr$ are \emph{bisimilar}, 
written $P \bisim Q$, if there exists a bisimulation $R$ such that 
$(P,Q) \in R$.

In Section~\ref{sect-sound-complete}, it is proved that $\bisim$ is a
congruence relation with respect to the operators of \STPA\REC.
Because of this, $\bisim$ induces a model of \STPA\REC.

The \emph{bisimulation} model of \STPA\REC\ is the quotient algebra of 
the term algebra over the signature of \STPA\REC\ modulo $\bisim$.

The additional rules for the maximal progress operators are given in 
Table~\ref{rules-maxpr}
\begin{table}[!t]
\caption{Additional rules for the maximal progress operators}
\label{rules-maxpr}
\begin{ruletbl}
\Rule
{\astp{x}{t,\sigma}{a}{x'},\; \nastp{x}{t,\sigma}{b}
 \mbox{for all $b \in \AAct$ with $a <_H b$}}
{\astp{\maxpr{H}(x)}{t,\sigma}{a}{\maxpr{H}(x')}}
\\
\Rule
{\astp{x}{t,\sigma}{a}{\!\term},\; \nastp{x}{t,\sigma}{b}
 \mbox{for all $b \in \AAct$ with $a <_H b$}}
{\astp{\maxpr{H}(x)}{t,\sigma}{a}{\!\term}}
\\
\Rule
{\istp{x}{t,\sigma}{t'},\; \nastp{x}{t,\sigma}{b}
 \mbox{for all $b \in \Union \set{\AAct(t'') \where t'' < t'}$ with
       $b \in H$}}
{\istp{\maxpr{H}(x)}{t,\sigma}{t'}}
\\
\Rule
{\astp{x}{t,\sigma}{a}{x'},\; \nastp{y}{t,\sigma}{b}
 \mbox{for all $b \in \AAct$ with $a <_H b$}}
{\astp{x \auxpr{H} y}{t,\sigma}{a}{\maxpr{H}(x')}}
\\
\Rule
{\astp{x}{t,\sigma}{a}{\!\term},\; \nastp{y}{t,\sigma}{b}
 \mbox{for all $b \in \AAct$ with $a <_H b$}}
{\astp{x \auxpr{H}y}{t,\sigma}{a}{\!\term}}
\\
\Rule
{\istp{x}{t,\sigma}{t'},\; \nastp{y}{t,\sigma}{b}
 \mbox{for all $b \in \Union \set{\AAct(t'') \where t'' < t'}$ with
       $b \in H$}}
{\istp{x \auxpr{H} y}{t,\sigma}{t'}}
\end{ruletbl}
\end{table}
and the additional rules for the recursion constants are given in 
Table~\ref{rules-recursion}.
\begin{table}[!t]
\caption{Additional rules for the recursion constants}
\label{rules-recursion}
\begin{ruletbl}
\Rule
{\astp{\rec{T}{E}}{t,\sigma}{a}{x'}\; [X \!=\! T \in E]}
{\astp{\rec{X}{E}}{t,\sigma}{a}{x'}}
\quad
\Rule
{\astp{\rec{T}{E}}{t,\sigma}{a}{\surd}\,\; [X \!=\! T \in E]}
{\astp{\rec{X}{E}}{t,\sigma}{a}{\surd}}
\\
\Rule
{\istp{\rec{T}{E}}{t,\sigma}{t'}\; [X \!=\! T \in E]}
{\istp{\rec{X}{E}}{t,\sigma}{t'}}
\end{ruletbl}
\end{table}

\section{Soundness and Completeness}
\label{sect-sound-complete}

In this section, soundness and (semi-)completeness results with respect 
to bisimimilarity for the axioms of \STPA\REC\ and \STPAt\REC\ are 
presented.
The results concerned are preceded by a congruence result for 
bisimilarity.

We have the following congruence result for bisimilarity.
\begin{lemma}[Congruence]
\label{lemma-congruence}
\sloppy
Bisimilarity based on the structural operation\-al semantics of 
\STPA\REC\ is a congruence with respect to the operators of 
\STPA\REC.
\end{lemma}
\begin{proof}
According to the definitions of a \emph{well-founded} rule and a rule in 
\emph{path format} in~\cite{BV93a}, all rules of the structural 
operational semantics of \STPA\REC\ are well-founded rules in path 
format. 
It follows by Theorem~5.4 of~\cite{BV93a} that bisimilarity based on the 
structural operational semantics of \STPA\REC\ is a congruence with 
respect to the operators of \STPA\REC.
\qed
\end{proof}

Years ago, in 2009, we devised an operational semantics for \STPA\REC\ 
consisting~of:
\begin{itemize}
\item
a binary relation 
${\step{\ell}}$ on $\Procr \x \Time \x \State$ 
for each $\ell \in \AAct$;
\item
a unary relation 
${\step{\ell}\!\term}$ on $\Procr \x \Time \x \State$ 
for each $\ell \in \AAct$;
\item
a unary relation ${\idle{\ell}}$ on $\Procr \x \Time \x \State$ 
for each $\ell \in \Time$
\end{itemize}
and a notion of bisimilarity that is an instance of the general notion 
of initially stateless bisimilarity from~\cite{MRG05a}.
The operational semantics was not in the format that would guarantee, 
according to Theorem~34 of~\cite{MRG05a}, that this bisimilarity 
relation is a congruence with respect to the operators of \STPA\REC, and 
we were unable to prove this otherwise.
It took many years before we realized that there exists a different 
operational semantics for \STPA\REC, which yields the same bisimilarity 
relation, but is in the path format of~\cite{BV93a}.

We have the following soundness result for \STPA\REC.
\begin{theorem}[Soundness]
\label{theorem-soundness}
For all $P,Q \in \Procr$, $\STPA\REC \Ent P = Q$ only if $P \bisim Q$.
\end{theorem}
\begin{proof}
Because ${\bisim}$ is a congruence with respect to all operators from 
the signature of \STPA\REC, it is sufficient to prove the validity of 
each axiom of \STPA\REC.

Below, we write $\csi(\eqn)$, where $\eqn$ is an equation of terms over 
the signature of \STPA\REC, for the set of all closed substitution 
instances of $\eqn$.
Moreover, we write $R_\id$ for the identity relation on $\Procr$.

For each axiom $\ax$ of \STPA\REC\ except the axiom 
$x \parc y = x \leftm y \altc y \leftm x$ and the instances of the axiom 
schema RSP, a bisimulation $R_\ax$ witnessing the validity of $\ax$ can 
be constructed as follows: 
\begin{ldispl}
R_\ax = \set{\tup{P,P'} \where P = P' \in \csi(\ax)} \union R_\id\;.
\end{ldispl}%
If $\ax$ is the axiom $x \parc y = x \leftm y \altc y \leftm x$, then
a bisimulation $R_\ax$ witnessing the validity of $\ax$ can be 
constructed as follows: 
\begin{ldispl}
\begin{array}[t]{@{}r@{\;}c@{\;}l@{}}
R_\ax  & = &
\set{\tup{P,P'} \where P = P' \in \csi(\ax)} \union R_\id \\ 
& \union &
\set{\tup{P,P'} \where P = P' \in \csi(x \parc y = y \parc x)}\;.
\end{array}
\end{ldispl}%
If $\ax$ is an instance 
$\set{X_i = T_i \where i \in I} \Implies 
 X_j = \rec{X_j}{\set{X_i = T_i \where i \in I}}$ 
($j \in I$) of~RSP, then a bisimulation $R_\ax$ witnessing the validity 
of $\ax$ can be constructed as follows:
\begin{ldispl}
R_\ax = 
\set{\tup{\vartheta(X_j),\rec{X_j}{\set{X_i = T_i \where i \in I}}} 
     \where 
\\ \phantom{R_\ax = {}\{}\, 
     j \in I \And \vartheta \in \Theta \And
     \AND_{i \in I} \vartheta(X_i) \bisim \vartheta(T_i)}
 \union R_\id\;,
\end{ldispl}%
where 
$\Theta$ is the set of all functions from $\cX$ to $\Procr$ and 
$\vartheta(T)$, where $\vartheta \in \Theta$ and $T$ is a term over the 
signature of \STPA, stands for $T$ with, for all $X \in \cX$, all 
occurrences of $X$ replaced by $\vartheta(X)$.

For each equational axiom $\ax$ of \STPA\REC, it is straightforward to 
check that the constructed relation $R_\ax$ is a bisimulation 
witnessing, for each closed substitution instance $P = P'$ of $\ax$, 
$P \bisim P'$.
For each conditional equational axiom $\ax$ of \STPA\REC, i.e.\ for 
each instance of RSP, it is straightforward to check that the 
constructed relation $R_\ax$ is a bisimulation witnessing, for 
each closed substitution instance 
$\set{P_i = P'_i \where i \in I} \Implies P = P'$ 
of $\ax$, $P \bisim P'$ if $P_i \bisim P'_i$ for each $i \in I$.
\qed
\end{proof}

We do not know whether the axioms of \STPA\REC\ are complete with 
respect to ${\bisim}$ for equations between terms from $\Procr$.
When applying the various methods developed to prove or disprove it, we 
keep getting stuck.
A major obstacle is that we cannot find a form in which all terms from 
$\Procr$ can be brought and in which the operational semantics is 
reasonably reflected.
However, we know that the axioms of \STPA\REC\ are complete in some 
degree, namely for equations between terms of the form 
$\state{C}{t}{\sigma}(P)$.
Below this semi-completeness result is proved. 
To do so, some additional definitions and lemmas are used.

Let $P,P' \in \Procr$.
Then $P$ is a \emph{summand} of $P'$, written $P \summand P'$, iff 
there exists a $P'' \in \Procr$ such that $P \altc P'' = P'$ or $P = P'$ 
is derivable from the axioms $x \altc y = y \altc x$ and
$(x \altc y) \altc z = x \altc (y \altc z)$ of \STPA.

The set $\SHProc$ of \emph{semi-head normal form process terms} and 
the auxiliary set $\SHProc'$ are the smallest subsets of $\Procr$ 
satisfying the following rules:
\begin{itemize}
\item
if $t \in \Time$, then $\adead{t}, \rdead{t} \in \SHProc'$;
\item
if $a \in \PAct \union \AAct$, then $a \in \SHProc'$;
\item
if $P \in \SHProc'$ and $Q \in \SHProc'$, then $P \tout Q \in \SHProc'$;
\item
if $P \in \SHProc'$, then $P \in \SHProc$;
\item
if $P \in \SHProc'$ and $Q \in \Procr$, then $P \seqc Q \in \SHProc$;
\item
if $P,Q \in \SHProc$, then $P \altc Q \in \SHProc$.
\end{itemize}
The set $\HProc$ of \emph{head normal form process terms} is the 
smallest subset of $\Procr$ satisfying the following rules:
\begin{itemize}
\pagebreak[2]
\item
if $t \in \Time$, then $\adead{t} \in \HProc$;
\item
if $a \in \AAct$, then $a \in \HProc$;
\item
if $a \in \AAct$ and $P \in \Procr$, then $a \seqc P \in \HProc$;
\item
if $P,Q \in \HProc$, then $P \altc Q \in \HProc$.
\end{itemize}
It is clear that $\HProc \subset \SHProc$.

We have the following result concerning $\SHProc$.
\begin{lemma}[Elimination]
\label{lemma-semi-hnf}
For each $P \in \Procr$, there exists a $Q \in \SHProc$ such that 
$\STPA\REC \Ent P = Q$.
\end{lemma}
\begin{proof}
This is straightforwardly proved by induction on the structure of $P$.
The cases where $P$ is of the form $\adead{t}$, $\rdead{t}$ or $a$
($a \in \PAct \union \AAct$) are trivial.
The case where $P$ is of the form $\rec{X}{E}$ follows immediately from
RDP and the easily proved claim that, for the unique term $T$ such that 
$X \!=\! T \in E$, $\rec{T}{E} \in \SHProc$.
The case where $P$ is of the form $P_1 \altc P_2$ follows immediately 
from the induction hypothesis. 
The case where $P$ is of the form $P_1 \parc P_2$ follows immediately
from the case where $P$ is of the form $P_1 \leftm P_2$.
Each of the remaining four cases follows immediately from the induction 
hypothesis and a claim that is easily proved by structural induction.
\qed
\end{proof}

It follows from the proof of Lemma~\ref{lemma-semi-hnf} that there 
exist $P \in \Procr$ for which there does not exist a $Q \in \HProc$ 
such that $\STPA\REC \Ent P = Q$.
In such cases, there are one or more occurrences of the time-out 
operator that cannot be eliminated.
This is due to the presence of potential receipts or the presence of 
both absolute timing and relative timing.
As mentioned earlier, the problem is that the point in time at which a 
potential receipt takes place is not fixed and that the initialization 
time needed to relate the two kinds of timing is not fixed.
However, the time-out operator can always be fully eliminated from terms 
in $\Procr$ that have the form $\state{C}{t}{\sigma}(P)$, provided $C$ 
includes all channels occurring in $P$.
\begin{lemma}[Elimination]
\label{lemma-hnf}
For all $C \subseteq \Chan$, $t \in \Time$, $\sigma \in \State$, and 
$P \in \Procr$ in which only channels from $C$ occur, there exists a 
$Q \in \HProc$ such that: 
\begin{itemize}
\item
$\STPA\REC \Ent \state{C}{t}{\sigma}(P) = Q$; 
\item
for all $a \in \AAct$ and $Q' \in \Procr$, $a \seqc Q' \summand Q$ only 
if there exist a $t' \in \Time$, $\sigma' \in \State$, and 
$P' \in \Procr$ in which only channels from $C$ occur such that 
$Q' \equiv \state{C}{t'}{\sigma'}(P')$.
\end{itemize}
\end{lemma}
\begin{proof}
By Lemma~\ref{lemma-semi-hnf}, it is sufficient to prove this 
theorem for all $P \in \SHProc$.
This is straightforwardly proved by induction on the structure of $P$.
The cases where $P$ is of the form $\adead{t}$, $\rdead{t}$ or $a$
($a \in \PAct \union \AAct$) are trivial.
The case where $P$ is of the form $P_1 \altc P_2$ follows immediately 
from the induction hypothesis. 
Each of the remaining two cases follows immediately from the induction 
hypothesis and a claim that is easily proved by structural induction.
\qed
\end{proof}

For $P \in \Procr$ for which there exists a $Q \in \HProc$ such that 
$\STPA\REC \Ent P = Q$, we write $\hnf(P)$ for a fixed but arbitrary 
$Q \in \HProc$ such that $\STPA\REC \Ent P = Q$.

Each closed term over the signature of \STPA\REC\ that is of the form
$\state{C}{t}{\sigma}(P)$ can be reduced to a linear recursive 
specification over the signature of \STPA. 

\begin{lemma}[Reduction]
\label{lemma-linear}
For all $C \subseteq \Chan$, $t \in \Time$, $\sigma \in \State$, and 
$P \in \Procr$ in which only channels from $C$ occur, there exists a 
linear recursive specification $E$ over the signature of \STPA\ and 
$X \in \vars(E)$ such that 
$\STPA\REC \Ent \state{C}{t}{\sigma}(P) = \rec{X}{E}$.
\end{lemma}
\begin{proof}
We approach this algorithmically.
In the construction of the linear recursive specification $E$, we keep
a set $F$ of recursion equations from $E$ that are already found and a
sequence $G$ of equations of the form $X_k = P_k$ with $P_k \in \HProc$ 
that still have to be transformed.
The algorithm has a finite or countably infinite number of stages.
In each stage, $F$ and $G$ are finite.
Initially, $F$ is empty and $G$ contains only the equation 
$X_0 = P_0$, where $P_0 \equiv \hnf(\state{C}{t}{\sigma}(P))$.
By Lemma~\ref{lemma-hnf}, $\hnf(\state{C}{t}{\sigma}(P))$ exists.

In each stage, we remove the first equation from $G$.
Assume that this equation is $X_k = P_k$. 
Assume that $P_k$ is $\Altc{i<n} a_i \seqc P'_i \altc \Altc{j<m} b_j$.
Then, we add the equation 
$X_k = \Altc{i<n} a_i \seqc X_{k+i+1} \altc  \Altc{j<m} b_j$,
where the $X_{k+i+1}$ are fresh variables, to the set $F$.
Moreover, for each $i < n$, we add the equation $X_{k+i+1} = P_{k+i+1}$,
where $P_{k+i+1} \equiv \hnf(P'_i)$, to the end of the sequence $G$.
By Lemma~\ref{lemma-hnf}, $\hnf(P'_i)$ exists.

Because $F$ grows monotonically, there exists a limit. 
That limit is the finite or countably infinite linear recursive 
specification $E$.
Every equation that is added to the finite sequence $G$, is also removed 
from it.
Therefore, the right-hand side of each equation from $E$ only contains
variables that also occur as the left-hand side of an equation from 
$E$.

Now, we want to use RSP to show that 
$\state{C}{t}{\sigma}(P) = \rec{X_0}{E}$.
The variables occurring in $E$ are $X_0, X_1, X_2, \ldots\;$.
For each $k$, the variable $X_k$ has been exactly once in $G$ as the 
left-hand side of an equation.
For each $k$, assume that  this equation is $X_k = P_k$.
To use RSP, we have to show for each $k$ that the equation
$X_k = \Altc{i<n} a_i \seqc X_{k+i+1} \altc  \Altc{j<m} b_j$ from $E$
with, for each $l$, all occurrences of $X_l$ replaced by $P_l$ is 
derivable from the axioms of \STPA\REC.
For each $k$, this follows from the construction.
\qed
\end{proof}

We have the following semi-completeness result for \STPA\REC.
\begin{theorem}[Semi-completeness]
\label{theorem-semi-complete}
For all 
$C,C' \in \Chan$, $t,t' \in \Time$, $\sigma,\sigma' \in \State$, 
$P \in \Procr$ in which only channels from $C$ occur, and
$P' \in \Procr$ in which only chan\-nels from $C'$ occur, 
$\STPA\REC \Ent \state{C}{t}{\sigma}(P) = \state{C'}{t'}{\sigma'}(P')$ 
if $\state{C}{t}{\sigma}(P) \bisim \state{C'}{t'}{\sigma'}(P')$.
\end{theorem}
\begin{proof}
By Theorem~\ref{theorem-soundness}, and Lemma~\ref{lemma-linear}, it 
suffices to prove that, for all linear recursive specifications $E$ and 
$E'$ with $X \in \vars(E)$ and $X' \in \vars(E')$, 
$\STPA\REC \Ent \rec{X}{E} = \rec{X'}{E'}$ if 
$\rec{X}{E} \bisim \rec{X'}{E'}$. 
This is proved in the same way as it is done for \ACP\REC\ in the proof 
of Theorem~4.4.1 from~\cite{Fok00}.
\qed
\end{proof}

In the case of \STPAt\REC, we have the following congruence result for 
bisimilarity.
\begin{lemma}[Congruence]
\label{lemma-congruence-maxpr}
\sloppy
Bisimilarity based on the structural operational semantics of 
\STPAt\REC\ is a congruence with respect to the operators of 
\STPAt\REC.
\end{lemma}
\begin{proof}
According to the definitions of a \emph{well-founded} rule and
a rule in \emph{panth format} in~\cite{Ver94b}, all rules of the 
structural operational semantics of \STPAt\REC\ are well-founded rules 
in panth format. 
Moreover, according to the definition of a \emph{stratified} set of 
rules in~\cite{Ver94b}, the set of rules of the structural operational 
semantics of \STPAt\REC\ is stratifiable.
It follows by Theorem~4.5 of~\cite{Ver94b} that bisimilarity based on 
the structural operational semantics of \STPAt\REC\ is a congruence with 
respect to the operators of \STPAt\REC.
\qed
\end{proof}

We have the following soundness result for \STPAt\REC.
\begin{theorem}[Soundness]
\label{theorem-soundness-maxpr}
For all closed terms $P$ and $Q$ over the signature of \STPAt\REC, 
$\STPAt\REC \Ent P = Q$ only if $P \bisim Q$.
\end{theorem}
\begin{proof}
Because ${\bisim}$ is a congruence with respect to all operators from 
the signature of \STPAt\REC, it is sufficient to prove the validity of 
each axiom of \STPAt\REC.
We use the same notation as in the proof of 
Theorem~\ref{theorem-soundness}.

For each axiom $\ax$ of \STPA\REC, a bisimulation $R_\ax$ witnessing the 
validity of $\ax$ can be constructed as in the proof of 
Theorem~\ref{theorem-soundness}. 
For each additional axiom $\ax$ of \STPAt\REC, a bisimulation $R_\ax$ 
witnessing the validity of $\ax$ can be constructed as for most axioms 
of \STPA\REC: 
\begin{ldispl}
R_\ax = \set{\tup{P,P'} \where P = P' \in \csi(\ax)} \union R_\id\;.
\end{ldispl}%

For each additional axiom $\ax$ of \STPAt\REC, it is straightforward to 
check that the constructed relation $R_\ax$ is a bisimulation 
witnessing, for each closed substitution instance $P = P'$ of $\ax$, 
$P \bisim P'$.
\qed
\end{proof}

We have the following semi-completeness result for \STPAt\REC.
\begin{theorem}[Semi-completeness]
\label{theorem-semi-complete-maxpr}
For all $H,H' \subseteq \AAct$,
$C,C' \in \Chan$, $t,t' \in \Time$, $\sigma,\sigma' \in \State$, 
$P \in \Procr$ in which only channels from $C$ occur, and
$P' \in \Procr$ in which only channels from $C'$ occur, 
$\STPAt\REC \Ent
 \maxpr{H}(\state{C}{t}{\sigma}(P)) =
 \maxpr{H'}(\state{C'}{t'}{\sigma'}(P'))$
if 
$\maxpr{H}(\state{C}{t}{\sigma}(P)) \bisim 
 \maxpr{H'}(\state{C'}{t'}{\sigma'}(P'))$.
\end{theorem}
\begin{proof}
It is easy to check that, for each $H \subseteq \AAct$, linear recursive 
specification $E$ over the signature of \STPA, and $X \in \vars(E)$, 
there exists a linear recursive specification $E'$ over the signature of 
\STPA\ and $X' \in \vars(E')$ such that 
$\STPAt\REC \Ent \maxpr{H}(\rec{X}{E}) = \rec{X'}{E'}$.
Moreover, we know from the proof of Theorem~\ref{theorem-semi-complete} 
that, for all linear recursive specifications $E$ and $E'$ over the 
signature of \STPA\ with $X \in \vars(E)$ and $X' \in \vars(E')$, 
$\STPAt\REC \Ent \rec{X}{E} = \rec{X'}{E'}$ if 
$\rec{X}{E} \bisim \rec{X'}{E'}$. 
The result follows immediately by Theorem~\ref{theorem-soundness-maxpr} 
and Lemma~\ref{lemma-linear}. 
\qed
\end{proof}

\section{Concluding Remarks}
\label{sect-conclusions}

In~\cite{BB93a,BM03b}, ACP-based process algebras for the timed 
behaviour of distributed systems with a known spatial distribution are 
presented in which the integration operator is needed to model 
asynchronous communication.
This operator is a variable-binding operator and therefore does not 
really fit in with an algebraic approach.
Moreover, a process algebra with this operator is not firmly founded in 
established meta-theory from the fields of universal algebra and 
structural operational semantics.
In this paper, we have presented an ACP-based process algebra for the 
timed behaviour of distributed systems with a known spatial distribution 
in which asynchronous communication can be modelled without the 
integration operator or another variable-binding operator.
The absolutely and relatively timed potential receive action constants 
along with special state operators obviate the need for a 
variable-binding operator.

In the setting of ACP, asynchronous communication has been studied 
previously.
The studies presented in~\cite{BKT85,BKP92a} abstract from aspects of 
space and time and the studies presented in~\cite{BB93a,BM03b} and the
current paper do not abstract from aspects of space and time.

The timed potential receive action constants introduced in the current 
paper provide a lower bound and an upper bound for the point in time at 
which a receive action can actually be performed.
In the setting of CCS~\cite{Mil80,Mil89}, an operator has been proposed 
in~\cite{Che92a} that provides a lower bound and an upper bound for 
the point in time at which an action can be performed.
That operator is in fact a limited integration operator. 
To the best of our knowledge, the CCS-based process algebra proposed
in~\cite{Che92a} has never been used to model asynchronous communication 
in distributed systems.

In~\cite{ST95a}, asynchronous communication in distributed systems is
studied in the setting of CCS.
That study has a very abstract view on the spatial distribution of a
distributed system.
Moreover, a lower bound and an upper bound for the point in time at 
which something can actually be received cannot be given in any way. 
In effect, these bounds are fixed at $0$ and $\infty$, respectively.
We could not find studies on asynchronous communication in distributed 
systems in the setting of CSP~\cite{BHR84,Hoa85}.

In this paper, the focus is on asynchronous communication in space-time.
However, \STPA\ can be easily extended with the action renaming and 
spatial replacement operators from~\cite{BB93a,BM03b}.
These operators facilitate dealing with a number of processes that 
differ only in the channels used for communication and dealing with 
processes that move in space.
Moreover, the state operators of \STPA\ can be easily adapted to deal 
uniformly with all transmission limitations due to blocking solid 
objects (as~in~\cite{BM03b}). 

\bibliographystyle{splncs04}
\bibliography{PA}

\begin{thebibliography}{10}
\providecommand{\url}[1]{\texttt{#1}}
\providecommand{\urlprefix}{URL }
\providecommand{\doi}[1]{https://doi.org/#1}

\bibitem{ABV94a}
Aceto, L., Bloom, B., Vaandrager, F.W.: Turning {SOS} rules into equations.
  Information and Computation  \textbf{111}(1),  1--52 (1994)
  \doi{10.1006/inco.1994.1040}

\bibitem{BB93a}
Baeten, J.C.M., Bergstra, J.A.: Real space process algebra. Formal Aspects of
  Computing  \textbf{5}(6),  481--529 (1993)
  \doi{10.1007/BF01211247}

\bibitem{BM02a}
Baeten, J.C.M., Middelburg, C.A.: Process Algebra with Timing, Monographs in
  Theoretical Computer Science, An EATCS Series, Springer-Verlag, Berlin (2002)
  \doi{10.1007/978-3-662-04995-2}.

\bibitem{BV93a}
Baeten, J.C.M., Verhoef, C.: A congruence theorem for structured operational
  semantics with predicates. In: Best, E. (ed.) CONCUR'93. Lecture Notes in
  Computer Science, vol.~715, pp. 477--492. Springer-Verlag (1993)
  \doi{10.1007/3-540-57208-2_33}

\bibitem{BW90}
Baeten, J.C.M., Weijland, W.P.: Process Algebra, Cambridge Tracts in
  Theoretical Computer Science, vol.~18. Cambridge University Press, Cambridge
  (1990)
  \doi{10.1017/CBO9780511624193}

\bibitem{BBP13a}
Bergstra, J.A., Bethke, I., Ponse, A.: Cancellation meadows: A generic basis
  theorem and some applications. Computer Journal  \textbf{56}(1),  3--14
  (2013)
  \doi{10.1093/comjnl/bxs028}

\bibitem{BBP15a}
Bergstra, J.A., Bethke, I., Ponse, A.: Equations for formally real meadows.
  Journal of Applied Logic  \textbf{13}(2B),  1--23 (2015)
  \doi{10.1016/j.jal.2015.01.004}

\bibitem{BK84b}
Bergstra, J.A., Klop, J.W.: Process algebra for synchronous communication.
  Information and Control  \textbf{60}(1--3),  109--137 (1984)
  \doi{10.1016/S0019-9958(84)80025-X}

\bibitem{BKT85}
Bergstra, J.A., Klop, J.W., Tucker, J.V.: Process algebra with asynchronous
  communication mechanisms. In: Brookes, S.D., Roscoe, A.W., Winskel, G. (eds.)
  Proceedings Seminar on Concurrency. Lecture Notes in Computer Science,
  vol.~197, pp. 76--95. Springer-Verlag (1985)
  \doi{10.1007/3-540-15670-4_4}

\bibitem{BM03b}
Bergstra, J.A., Middelburg, C.A.: Located actions in process algebra with
  timing. Fundamenta Informaticae  \textbf{61}(3--4),  183--211 (2004)
  \link{content.iospress.com/articles/fundamenta-informaticae/fi61-3-4-01}

\bibitem{BKP92a}
de~Boer, F.S., Klop, J.W., Palamidessi, C.: Asynchronous communication in
  process algebra. In: LICS '92. pp. 137--147. IEEE (1992)
  \doi{10.1109/LICS.1992.185528}

\bibitem{BHR84}
Brookes, S.D., Hoare, C.A.R., Roscoe, A.W.: A theory of communicating
  sequential processes. Journal of the ACM  \textbf{31}(3),  560--599 (1984)
  \doi{10.1145/828.833}

\bibitem{Che92a}
Chen, L.: An interleaving model for real-time systems. In: Nerode, A.,
  Taitslin, M. (eds.) Symposium on Logical Foundations of Computer Science.
  Lecture Notes in Computer Science, vol.~620, pp. 81--92. Springer-Verlag
  (1992)
  \doi{10.1007/BFb0023865}

\bibitem{Fok00}
Fokkink, W.J.: Introduction to Process Algebra. Texts in Theoretical Computer
  Science, An EATCS Series, Springer-Verlag, Berlin (2000)
  \doi{10.1007/978-3-662-04293-9}

\bibitem{Gog21a}
Goguen, J.A.: Theorem proving and algebra. {\tt arXiv:2101.02690 [cs.LO]}
  (January 2021) \arXiv{2101.02690}

\bibitem{Hoa85}
Hoare, C.A.R.: Communicating Sequential Processes. Prentice-Hall, Englewood
  Cliffs (1985)

\bibitem{Mid02b}
Middelburg, C.A.: Process algebra with nonstandard timing. Fundamenta
  Informaticae  \textbf{53}(1),  55--77 (2002)
  \link{content.iospress.com/articles/fundamenta-informaticae/fi53-1-03}

\bibitem{Mil80}
Milner, R.: A Calculus of Communicating Systems, Lecture Notes in Computer
  Science, vol.~92. Springer-Verlag, Berlin (1980)
  \doi{10.1007/3-540-10235-3}

\bibitem{Mil89}
Milner, R.: Communication and Concurrency. Prentice-Hall, Englewood Cliffs
  (1989)

\bibitem{MRG05a}
Mousavi, M.R., Reniers, M.A., Groote, J.F.: Notions of bisimulation and
  congruence formats for {SOS} with data. Information and Computation
  \textbf{200},  107--147 (2005)
  \doi{10.1016/j.ic.2005.03.002}

\bibitem{ST95a}
Satoh, I., Tokoro, M.: A formalism for remotely interacting processes. In: Ito,
  T., Yonezawa, A. (eds.) TPPP '94. Lecture Notes in Computer Science,
  vol.~907, pp. 216--228. Springer-Verlag (1995)
  \doi{10.1007/BFb0026571}

\bibitem{Ver94b}
Verhoef, C.: A congruence theorem for structured operational semantics with
  predicates and negative premises. In: Jonsson, B., Parrow, J. (eds.)
  CONCUR '94. Lecture Notes in Computer Science, vol.~836, pp. 433--448.
  Springer-Verlag (1994)
  \doi{10.1007/978-3-540-48654-1_32}

\end{thebibliography}

% \par \mbox{} \par \vfill \par \noindent DRAFT of \today

%%%%%%%%%%%%%%%%%%%%%%%%%%%%%%%%%%%%%%%%%%%%%%%%%%%%%%%%%%%%%%%%%%%%%%%%

\begin{comment}
Owing to the presence of absolute timed actions, a time-out operator is 
introduced to axiomatize parallel composition of processes.
\end{comment}

\begin{comment}

In~\cite{Lut02a}, a serious attempt is made to fit in a variable-binding
operator related to the integration operator with an algebraic approach.

In~\cite{BB95d}, where essentially the same replacement has been made,
the replacing operator is called the left strong choice operator.

There is no problem in adding constants for all meadow elements.

Moreover, we use the notation $\aprcva{c}{d}{t}{\xi}$ for 
$\aprcv{c}{d}{t}{t}{\xi}$ and the notation $\rprcva{c}{d}{t}{\xi}$ for
$\rprcv{c}{d}{t}{0}{\xi}$.

This may be related to the fact that in CSP equality corresponds to 
failure equivalence instead of bisimilarity.

\end{comment}

\begin{comment}
An asynchronous version of the PAR (Positive Acknowledgement with 
Retransmission) protocol similar to the simplest asynchronous version of 
the PAR protocol specified in~\cite{BM03b} can be easily specified in 
\STPAt.
It mainly involves replacing terms of the forms
$\int_{v \in [t,t')} \sigma^v(c{\downarrow}d(\xi))$ and
$\int_{v \in [t,t')} \Altc{d \in D} \sigma^v(c{\downarrow}d(\xi))$
by terms of the forms $\rprcv{c}{d}{t}{t'}{\xi}$ and 
$\Altc{d \in D} \rprcv{c}{d}{t}{t'}{\xi}$, respectively.
Versions similar to the other versions specified in~\cite{BM03b} need 
the above-mentioned extensions of \STPAt. 
\end{comment}

\end{document}